\documentclass[prb,twocolumn,showpacs,preprintnumbers,amsmath,amssymb]{revtex4}
\usepackage{graphicx}
\usepackage{dcolumn}
\usepackage{bm}

\begin{document}

\title{Structure and optical properties of high light output halide
scintillators}

\author{David J. Singh}

\affiliation{Materials Science and Technology Division
and Center for Radiation Detection Materials and Systems,
Oak Ridge National Laboratory, Oak Ridge, Tennessee 37831-6114}

\date{\today}

\begin{abstract}
Structural and optical properties of several high light output
halide scintillators and closely related materials are presented
based on first principles calculations. 
The optical properties are based on the Engel-Vosko generalized gradient
approximation and the recently developed density functional
of Tran and Blaha.
The materials investigated are BaBr$_2$, BaIBr, BaCl$_2$, BaF$_2$,
BaI$_2$, BiI$_3$, CaI$_2$, Cs$_2$LiYCl$_6$, CsBa$_2$Br$_5$, CsBa$_2$I$_5$,
K$_2$LaBr$_5$, K$_2$LaCl$_5$,K$_2$LaI$_5$, LaBr$_3$, LaCl$_3$,
SrBr$_2$, and YI$_3$. For comparison results are
presented for the oxide CdWO$_4$.
We find that the Tran Blaha functional gives greatly improved band
gaps and optical properties in this class of materials. Furthermore,
we find that unlike CdWO$_4$, most of these halides are highly isotropic
from an optical point of view even though in many cases
the crystal structures and other properties are not. This general result
is rationalized in terms of halide chemistry. Implications for the
development of ceramic halide scintillators are discussed.
\end{abstract}


\maketitle

\section{introduction}

Scintillators are widely used in radiation detection applications including
medical imaging, oil well drilling, nuclear security and high
energy physics experiments. These materials
function by emitting light when excited by ionizing radiation, such as
gamma rays. This light is then coupled to a photomultiplier or
other light detector to produce electrical signals.
The performance of scintillators is characterized by their light
output, normally given in terms of photons per MeV of excitation energy,
proportionality (how linear the light output is as a 
function of excitation energy), density (related to stopping power),
decay time and energy resolution. \cite{knoll}

Scintillator performance depends on energy transport between
the energy absorption events and the scintillation centers,
suppression of non-radiative recombination channels and fast
efficient light emission from the scintillation center.
These scintillation centers can be intrinsic, via a mechanism
such as decay of a self trapped exciton, or, as is commonly the case,
at an activation center (e.g. Ce$^{3+}$ ions 
substituting for La${^3+}$ in LaBr$_3$) where electron hole pairs 
generated by the radiation recombine.
These processes are sensitive to the details of the material and in the
case of materials with activators the environment
of the activator ions.
Halides generally have soft lattices that disfavor
non-radiative recombination and can typically incorporate rare earth and
other activator ions.
More importantly, halide chemistry is very rich, with a wide variety of
crystal structures that provide different bonding topologies for energy
transfer and various environments for activator ions.
\cite{pauling,meyer}
In fact, one of the best known scintillators is the halide,
NaI activated with Tl$^{+}$.
This material has high light output, but is slow, non-proportional
and has poor energy resolution. \cite{dehaas}
The finding that Ce$^{3+}$ activated LaBr$_3$
is a very high light output proportional scintillator
with energy resolution better than 3\% at 662 KeV \cite{loef-labr3} has
led to renewed interest in halides as scintillator hosts, especially
for spectroscopic gamma ray detection.
This interest has resulted in
the discovery of several other interesting halide
scintillators,
including heavily Eu$^{2}$ activated SrI$_2$, which is very proportional
and has a light output exceeding that of LaBr$_3$, \cite{cherepy} and
Ce$^{3+}$ activated YI$_3$, which is another very
high light output material. \cite{glodo-yi3}

Full characterization of the optical properties of scintillators
is very useful both from the point of view of improving the design
of systems as regards light coupling and also importantly
in selecting candidate
materials for ceramic scintillators.
However, full optical characterization of these halides is complicated
by sensitivity to moisture and other experimental difficulties.
As a result only limited data is available.

The key requirements for a ceramic scintillator are sinterability
and optical isotropy. This latter requirement comes from the
need to avoid light scattering at misoriented grain boundaries.
It is commonly thought that because of
this requirement cubic material is needed for a true
transparent ceramic. However, radiation detection is not an optical
imaging application. Therefore weak scattering and image distortion
are not as detrimental as they would be in an optical application
such as a ceramic lens.
In fact ceramic scintillators based on monoclinic Lu$_2$SiO$_5$ (LSO)
have been demonstrated. \cite{lempicki,wisniewski}
We showed in previous work that the
high light output halide scintillator, SrI$_2$ is in fact very nearly
optically isotropic in spite of its
orthorhombic ($Pbca$) crystal 
structure. \cite{singh-sri2}
This was an expected result considering the strongly orthorhombic
lattice.
It is of interest to determine whether this is the case for other
high light output halide scintillators.
Here we present a consistent set of first principles data for
structural and optical properties of a number of high light
output halide scintillators and closely related materials.
We find, quite unexpectedly, that the halides we investigate
have remarkably little optical anisotropy, even though a number
of them are very anisotropic from other points of view.

\section{approach}

The density functional calculations reported here were performed
using the general potential linearized augmented planewave (LAPW) method
as implemented in the WIEN2K code. \cite{wien}
We used well converged Brillouin zone samples and basis sets, with the
standard LAPW augmentation plus local orbitals. \cite{singh-lo}
Relativity was included at the scalar relativistic level except for
BiI$_3$, where spin-orbit was included for the electronic structure.

The crystal structure plays a fundamental role in determining the
electronic and optical properties of a material.
We began our calculations by fully relaxing all free internal
atomic positions consistent with the crystal symmetry for each
material. The lattice parameters were held fixed at their experimental
values, which are no doubt more precise than the values that can
be obtained using density functional calculations.
This relaxation was done using the generalized gradient approximation (GGA)
of Perdew, Burke and Ernzerhof (PBE). \cite{pbe}.

The PBE functional, like other standard generalized gradient approximations,
is based on the total
energy in terms of the coupling constant averaged exchange correlation
hole and is designed to reproduce the total energy.
\cite{gga}
While these functionals are very useful in obtaining structures and
other properties related to total energies,
they underestimate, often strongly, the band gaps of most semiconductors
and simple insulators.
Accordingly, for the optical properties we use two other functionals. The
first is the Engel-Vosko GGA (EV). \cite{ev}
This functional was designed to reproduce the exchange-correlation potential
rather than the total energy, and gives improved band gaps.
\cite{dufek}
We used it in our prior studies
on transport in PbTe and optical properties
of SrI$_2$ and Bi$_4$Ge$_3$O$_{12}$ scintillators.
\cite{singh-pbte,singh-sri2,jellison-bgo}
The second is the semi-local functional of Tran and Blaha
(TB-mBJ). \cite{mbj}
This relatively recent
functional is more sophisticated than the Engel-Vosko GGA and
has been shown to give very much improved band gaps for a variety of
semiconductors and insulators.
We find that where comparison with experiment is possible
this functional also gives very much improved band
gaps as well as optical properties for these halides.
As such, we focus on the results obtained using the TB-mBJ functional.
The electronic structures were calculated with these two functionals
based on the relaxed crystal structures from the PBE calculations.
Optical properties were then obtained using the dipole matrix elements
with the WIEN2K optical package. \cite{wien}
A 0.1 eV broadening was applied to the spectra.

\section{materials and structures}

We begin with the calculated
structural parameters and brief introductions to the materials that we study.

BaF$_2$ is a very well characterized material.
It has been applied as a scintillator, both in pure form
and with activation.
\cite{laval,woody,visser,wojtowicz,dinca}
In pure form it has a
very fast component ($\sim$0.8 ns), as well as a slow component,
which can be at least partially suppressed by La doping.
Fast response is of importance in applications with very high count
rates or where timing is critical.
It occurs in
the cubic ($Fm\bar{3}m$) CaF$_2$ structure,
Ba on Wyckoff site $4a$ (0,0,0) and F on $8c$ (1/4,1/4,1/4).
In our calculations we used the experimental
lattice parameter,\cite{swanson} $a$=6.2001 \AA.

Several of the alkaline earth di-halides, $AeX_2$,
are high light output proportional scintillators when activated with
Eu$^{2+}$.
\cite{cherepy,hofstadter-cai2,hofstadter,jestin,selling}
As with BaF$_2$, activation with Ce$^{3+}$ is also possible in some cases.
\cite{selling-ce}
Undoped BaCl$_2$ has a very fast response, with a short component
lifetime of 1.6 ns. \cite{koshimizu}
SrI$_2$:Eu is equal to or superior to LaBr$_3$:Ce$^{3+}$
as regards light output and proportionality, although it
suffers from a slower response. \cite{cherepy}
Even though the material is orthorhombic, we found it to be optically
nearly isotropic. \cite{singh-sri2}
CaI$_2$:Eu$^{2+}$ is another material with very high light output,
$\sim100,000$ photons/MeV.
\cite{hofstadter,cherepy}
BaIBr:Eu$^{2+}$ crystals have been shown to have a light output of
81,000$\pm$3,000 photons/MeV with a 662 KeV energy resolution better
than 5\%.
\cite{bourret-babri}
Regarding structure,
BaCl$_2$, BaBr$_2$, BaI$_2$ and BaIBr occur in an orthorhombic $Pnma$,
PbCl$_2$ type structure with four formula units per cell.
The structural parameters as determined from relaxation are given
in Table \ref{struct-bacl2}. SrBr$_2$ has a large tetragonal
unit cell ($P4/nz$) with ten formula units per cell. \cite{smeggil}
The calculated internal parameters are as given in Table \ref{struct-srbr2}.
CaI$_2$ is hexagonal ($P\bar{3}m1$), with
Ca on site $1a$ (0,0,0) and I on $2d$ (1/2,2/3,$z$).
In our optical calculations we used the experimental
lattice parameters, $a$=4.49 \AA, $c$=6.975 \AA,
with the calculated $z$=0.2535 from
total energy minimization with the
PBE GGA (the reported experimental value is 0.25).
\cite{blum}

LaCl$_3$ and LaBr$_3$ activated with Ce$^{3+}$ are very high light
output, proportional scintillators with excellent energy resolution.
\cite{loef-labr3}
They are hexagonal ($P6_3/m$) with the UCl$_3$ structure
type. There are two formula units per cell, with La on site
$2c$ (1/3,2/3,1/4) and the halogen on $6h$ ($u$,$v$,1/4).
We used the experimental lattice parameters, $a$=7.4779 \AA, $c$=4.3745 \AA,
for LaCl$_3$ (Ref. \onlinecite{morosin})
and $a$=7.9648 \AA, $c$=4.5119 \AA, for LaBr$_3$ (Ref. \onlinecite{kraemer}).
The calculated internal parameters were
$u$=0.9145, $v$=0.6132 for LaCl$_3$
and $u$=0.9144, $v$=0.6159 for LaBr$_3$.

Cs$_2$LiYCl$_6$ is a member of the elpasolite family.
The elpasolites are cubic $Fm\bar{3}m$ halides, with general
formula $A1_2A2RX_6$, with $A1$ and $A2$ alkali metals,
$X$ a halogen and $R$ a 
rare earth or other trivalent element.
$A1$ is on site $8c$ (1/4,1/4,1/4),
$A2$ is on $4b$ (1/2,1/2,1/2), $R$ is on $4a$ (0,0,0)
and $X$ is on $24e$ ($u$,0,0).
A large number of such compounds are known. \cite{meyer}
However, only a fraction have been studied as potential 
scintillator materials.
\cite{eijk,loef,birowosuto,biro2,bessiere,grimm,higgins}
Cs$_2$LiYCl$_6$ is of particular interest because
of its high Li content, which makes it useful as a neutron detector.
We used the experimental lattice parameter for Cs$_2$LiYCl$_6$
from Reber and co-workers, \cite{reber} $a$=10.4857 \AA,
with the calculated internal parameter $u$=0.2517
(an experimental value of 0.25046 was reported by Reber and co-workers).

The scintillation properties of Ce$^{3+}$ activated
K$_2$LaCl$_5$, K$_2$LaBr$_5$ and K$_2$LaI$_5$ were
investigated by van Loef and co-workers. \cite{loef-215a,loef-215}
These isostructural orthorhombic compounds showed high light output,
up to 55,000 photons / MeV (for K$_2$LaI$_5$:Ce$^{3+}$).
The iodide also showed a reasonable decay time of 24$\pm$5 ns and
662 KeV energy resolution of 4.5$\pm$0.5\%,
although use of this scintillator is complicated because of self-activity
associated with K. Our calculated structural properties are give in
Table \ref{struct-k2lacl5}.

\begin{table}
\caption{Structural properties of
orthorhombic $Pnma$ PbCl$_2$ type, BaCl$_2$, BaBr$_2$, BaI$_2$ and BaIBr.
The lattice parameters are from experiment
(Refs. \onlinecite{sahl}, \onlinecite{brackett} and \onlinecite{frit}),
while the internal parameters are from total energy minimization using
the PBE functional. For BaIBr, $X1$ is I and $X2$ is Br. All atoms are
on Wyckoff site 4$c$ with $y$=0.25.
}
\begin{tabular}{|l|c|c|c|c|}
\hline
~~~~~~~~~~~ & ~~BaCl$_2$~~ & ~~BaBr$_2$~~ & ~~~BaI$_2$~~~ & ~~BaIBr~~  \\
\hline
$a$ (\AA) & 7.878 & 8.276 &  8.922 & 8.6~~ \\
$b$ (\AA) & 4.731 & 4.956 &  5.304 & 5.12~ \\
$c$ (\AA) & 9.415 & 9.919 & 10.695 & 10.31~ \\
Ba $x$    & 0.2481 & 0.2470 & 0.2466  & 0.2302  \\
Ba $z$    & 0.3838 & 0.1163 & 0.1166  & 0.1228 \\
$X$1 $x$  & 0.1423 & 0.6426 & 0.6424  & 0.9722 \\
$X$1 $z$  & 0.0708 & 0.0714 & 0.0721  & 0.6701 \\
$X$2 $x$  & 0.5274 & 0.5266 & 0.5246  & 0.3538 \\
$X$2 $z$  & 0.8313 & 0.6686 & 0.6665  & 0.4325 \\
\hline
\end{tabular}
\label{struct-bacl2}
\end{table}

\begin{table}
\caption{Internal structural coordinates of tetragonal
$P4/nz$ (No. 85) SrBr$_2$ as determined by total energy minimization.
The lattice parameters were fixed at their experimental
values of $a$=11.630 \AA, and $c$=7.146 \AA. (Ref. \onlinecite{smeggil})
}
\begin{tabular}{|l|cccccc|}
\hline
~~~~~~ & ~~~Sr1~~~ & ~~~Sr2~~~ & ~~~Br1~~~ & ~~~Br2~~~ & ~~~Br3~~~ & ~~~Br4~~~ \\
\hline
$x$  & 0.4141 & 0.7500 & 0.5402 & 0.5426 & 0.2500 & 0.2500 \\
$y$  & 0.6035 & 0.7500 & 0.6527 & 0.8386 & 0.7500 & 0.7500 \\
$z$  & 0.7520 & 0.1487 & 0.3761 & 0.9004 & 0.0000 & 0.5000 \\
\hline
\end{tabular}
\label{struct-srbr2}
\end{table}

\begin{table}
\caption{Structural properties of
orthorhombic $Pnma$ K$_2$La$X_5$, $X$=Cl,Br,I.
The lattice parameters are from experiment
(Refs. \onlinecite{meyer-k2lacl5} and \onlinecite{meyer-215}),
while the internal parameters are from total energy minimization using
the PBE functional. The $y$ coordinate is not listed for the atoms that
are on Wyckoff site $4c$ for which $y$=0.25.}
\begin{tabular}{|l|c|c|c|}
\hline
~~~~~~~~~~~ & ~~K$_2$LaCl$_5$~~ & ~~K$_2$LaBr$_5$~~ & ~~K$_2$LaI$_5$~~  \\
\hline
$a$ (\AA) & 12.742 & 13.36 & 14.332  \\
$b$ (\AA) & 8.868 & 9.272 & 9.912  \\
$c$ (\AA) & 8.022 & 8.462 & 9.132  \\
K $x$     & 0.3279 & 0.3277 & 0.3264 \\
K $y$     & 0.9952 & 0.5049 & 0.5035 \\
K $z$     & 0.0475 & 0.4465 & 0.4419 \\
La $x$    & 0.0066 & 0.0070 & 0.0064 \\
% La $y$    & 0.2500 & 0.2500 & 0.2500 \\
La $z$    & 0.0774 & 0.4202 & 0.4182 \\
$X$1 $x$  & 0.1818 & 0.0056 & 0.0053 \\
% $X$1 $y$  & 0.2500 & 0.2500 & 0.2500 \\
$X$1 $z$  & 0.8644 & 0.0646 & 0.0609 \\
$X$2 $x$  & 0.0065 & 0.1841 & 0.1847 \\
% $X$2 $y$  & 0.2500 & 0.2500 & 0.2500 \\
$X$2 $z$  & 0.4311 & 0.6329 & 0.6327 \\
$X$3 $x$  & 0.2912 & 0.2898 & 0.2879 \\
% $X$3 $y$  & 0.2500 & 0.2500 & 0.2500 \\
$X$3 $z$  & 0.3298 & 0.1674 & 0.1654 \\
$X$4 $x$  & 0.0816 & 0.0799 & 0.0792 \\
$X$4 $y$  & 0.5435 & 0.5459 & 0.5471 \\
$X$4 $z$  & 0.1659 & 0.3298 & 0.3282 \\
\hline
\end{tabular}
\label{struct-k2lacl5}
\end{table}

\begin{table}
\caption{Structural properties of
monoclinic $P2_1/c$ CsBa$_2X_5$, $X$=Br,I.
The lattice parameters are from experiment
(Refs. \onlinecite{schilling-csba2br5} and \onlinecite{bourret-csba2i5}),
while the internal parameters are from total energy minimization using
the PBE functional.}
\begin{tabular}{|l|ccc|ccc|}
\hline
~~~~~~~~~~~ & &  CsBa$_2$Br$_5$ & & & CsBa$_2$I$_5$ &  \\
\hline
$a,b,c$ (\AA) & 13.816 & 9.987 & 8.665 & 14.637 & 10.541 & 9.256 \\
$\gamma$ & & & 90.2 & & & 90.194 \\
Cs   & 0.1665 & 0.9897 & 0.5586 & 0.1675 & 0.9911 & 0.5594 \\
Ba1  & 0.1768 & 0.5002 & 0.5300 & 0.1782 & 0.4999 & 0.5332 \\
Ba2  & 0.4921 & 0.2511 & 0.5704 & 0.4913 & 0.2500 & 0.5772 \\
$X$1 & 0.4211 & 0.9614 & 0.6782 & 0.4205 & 0.9562 & 0.6810 \\
$X$2 & 0.0980 & 0.4602 & 0.1643 & 0.0956 & 0.4565 & 0.1669 \\
$X$3 & 0.4980 & 0.2849 & 0.9535 & 0.4985 & 0.2813 & 0.9592 \\
$X$4 & 0.3122 & 0.2868 & 0.3356 & 0.3122 & 0.2819 & 0.3375 \\
$X$5 & 0.2192 & 0.2732 & 0.8099 & 0.2216 & 0.2694 & 0.8110 \\
\hline
\end{tabular}
\label{struct-csba2br5}
\end{table}

CsBa$_2$Br$_5$ and CsBa$_2$I$_5$ occur in a monoclinic $P2_1/c$
(No. 14) structure.
\cite{schilling-csba2br5,bourret-csba2i5}
Eu$^{2+}$ doped CsBa$_2$I$_5$ was reported to have a light yield of
approximately 97,000 photons per MeV and to be
less hygroscopic than LaBr$_3$ or SrI$_2$. \cite{bourret-csba2i5}
The calculated structural parameters are as given in
Table \ref{struct-csba2br5}.

YI$_3$ and BiI$_3$ are related materials. Ce$^{3+}$ activated
YI$_3$ was recently reported to have a very high light yield of almost
100,000 photons per MeV. \cite{glodo-yi3,loef-yi3}
The related compound BiI$_3$ is of interest mainly as a
semiconductor for radiation detection,
\cite{nason}
rather than as a scintillator.
We include it here for comparison with YI$_3$. Unlike YI$_3$, BiI$_3$
has substantial covalency between Bi $6p$ and halogen $p$ states,
which leads to enhanced Born charges and may be expected to affect
the optical properties. \cite{du-2010}
These compounds occur in a rhombohedral $R\bar{3}$ structure,
with the cations on site $6c$ (0,0,$z$) and I on $18f$ ($u$,$v$,$w$).
With the hexagonal setting, the experimental lattice parameters are
$a$=7.4864 \AA, $c$=20.880 \AA, for YI$_3$ (Ref. \onlinecite{jongen}) and
$a$=7.516 \AA, $c$=20.7180 \AA, for BiI$_3$ (Ref. \onlinecite{trotter}).
The calculated internal coordinates are
$z$=0.1664, $u$=0.6568, $v$=0.9993, $w$=0.4156, for YI$_3$ and
$z$=0.1664, $u$=0.0022, $v$=0.6685, $w$=0.4123, for BiI$_3$.

\section{Electronic Structure and Optical Properties}

As mentioned, three functionals were used in the present study,
the PBE GGA for the structural properties, and the Engel-Vosko
and TB-mBJ functionals for the optical properties. The recently
developed TB-mBJ has been shown to give much more accurate band
gaps than other semilocal functionals in a wide variety of
materials. \cite{mbj} As such, we emphasize results obtained using
that functional. For comparison, the fundamental band gaps
of the compounds studied are listed in Table \ref{gaps}.
The EV gaps are invariably intermediate between those of the standard
PBE and the TB-mBJ functionals with the exception of the La containing
compounds. In those compounds, the $f$-resonance is found inside the
insulating gap for all three functionals. The position of the $f$-resonance
relative to the valence band edge depends on the compound,
but for a given material is practically
unchanged between the different functionals. For these compounds we
show both the gap between the valence band maximum and the bottom of the
$f$-bands making up the resonance, as well as the gap between the
valence bands and the non-$f$ conduction bands.
Experimental band gaps are unavailable for most of these materials.

\begin{figure}
\includegraphics[width=\columnwidth]{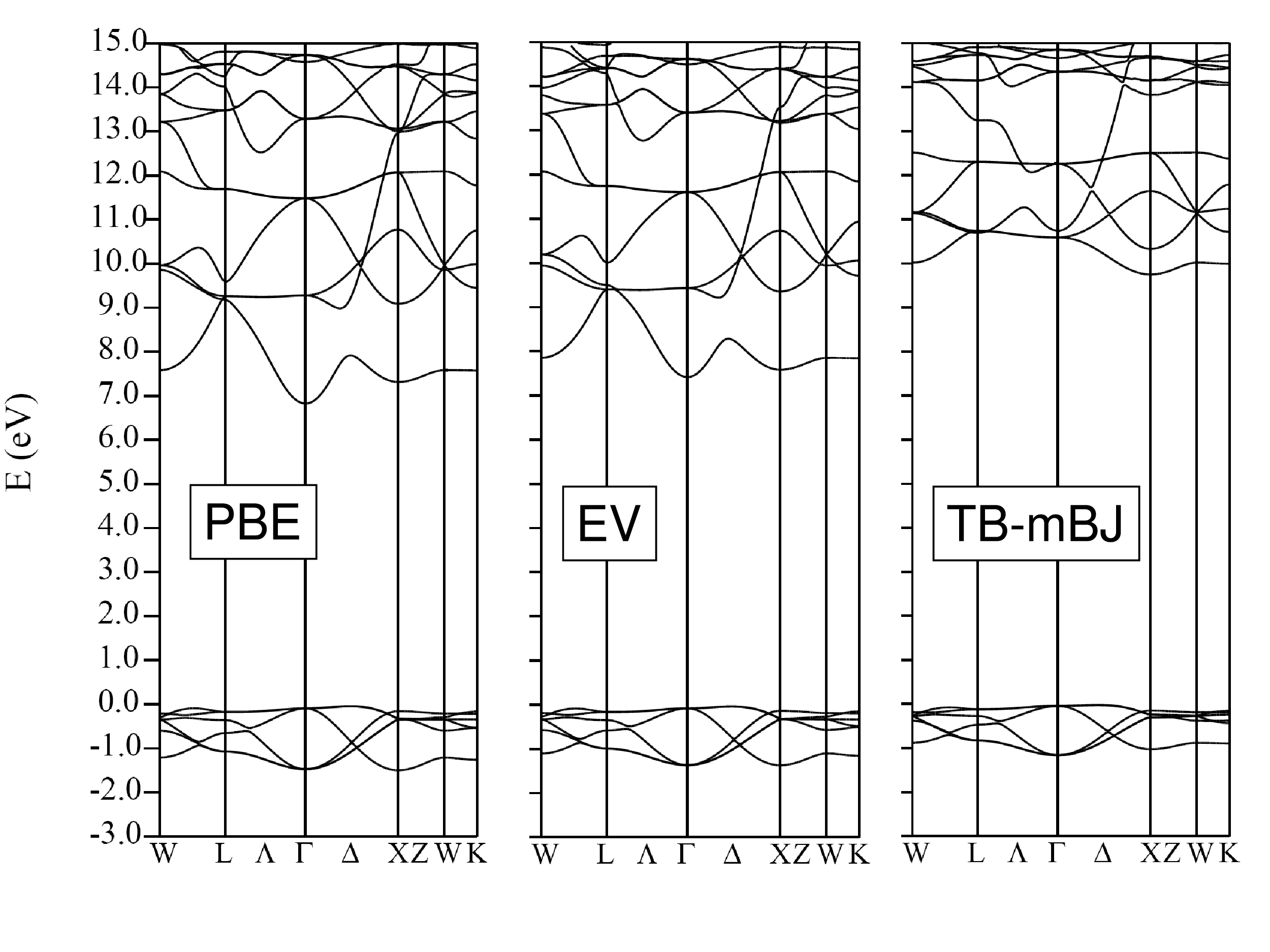}
\caption{Calculated band structure of BaF$_2$ with the three functionals.}
\label{BaF2-bands}
\end{figure}

\begin{figure}
\includegraphics[height=0.9\columnwidth,angle=270]{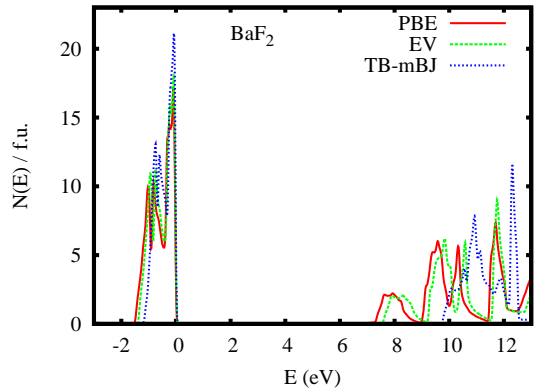}
\caption{(color online)
Electronic density of states of BaF$_2$ as calculated with the PBE,
EV and TB-mBJ functionals.}
\label{BaF2-dos}
\end{figure}

We start with BaF$_2$, which as mentioned is a relatively well
characterized material. The calculated band structure with the
PBE, EV and TB-mBJ functionals is as shown in Fig. \ref{BaF2-bands},
while the corresponding electronic density of states (DOS) is given
in Fig. \ref{BaF2-dos}.
The PBE band structure has a direct gap at $\Gamma$ that is
clearly underestimated with respect to experiment.
\cite{rubloff}
The valence bands, of F $p$ character, are narrow with a width of less
that 2 eV, while the metal derived conduction bands are more dispersive.
The EV band structure is similar to the
PBE result but with a somewhat larger band gap.
The mBJ band structure has a considerably enhanced gap of 9.8 eV.
Furthermore, the conduction bands are distorted relative to the PBE
result, with the conduction band minimum now at the $X$ point, although
it should be noted that the distinction between direct and indirect is
not so important for the properties of this compound because of the very
flat valence bands.

\begin{table}
\caption{Calculated fundamental band gaps in eV for the various
compounds. For La containing compounds we list both the
band gap from the valence band edge to the La $4f$-resonance
(denoted $f$)
and to the conduction bands excluding the resonance
(denoted cb, see text). Spin orbit is included for BiI$_3$.
}
\begin{tabular}{|l|ccc|}
\hline
~~~~~~~~~~~ & ~~~~PBE~~~~ & ~~~~EV~~~~ & ~~~TB-mBJ~~~ \\
\hline
BaF$_2$         & 6.87 & 7.46 & 9.77 \\
BaCl$_2$        & 5.23 & 5.66 & 6.45 \\
BaBr$_2$        & 4.45 & 4.85 & 5.39 \\
BaIBr           & 3.79 & 4.13 & 4.45 \\
BaI$_2$         & 3.50 & 3.87 & 4.14 \\
CaI$_2$         & 3.49 & 3.92 & 4.26 \\
SrBr$_2$        & 4.99 & 5.25 & 5.91 \\
YI$_3$          & 2.81 & 3.02 & 3.31 \\
BiI$_3$         & 1.43 & 1.67 & 1.82 \\
Cs$_2$LiYCl$_6$ & 4.90 & 5.30 & 5.97 \\
CsBa$_2$Br$_5$  & 4.53 & 5.13 & 5.77 \\
CsBa$_2$I$_5$   & 3.69 & 4.14 & 4.49 \\
K$_2$LaCl$_5 f$ & 4.46 & 4.46 & 4.59 \\
K$_2$LaCl$_5$ cb & 5.23 & 5.49 & 6.32 \\
K$_2$LaBr$_5 f$ & 3.69 & 3.76 & 3.74 \\
K$_2$LaBr$_5$ cb & 4.33 & 4.66 & 5.26 \\
K$_2$LaI$_5 f$  &  2.79 & 2.95 & 2.80 \\
K$_2$LaI$_5$ cb  &  ~3.38$^*$ & 3.61 & 3.99 \\
LaCl$_3 f$   & 3.97 & 4.04 & 4.15 \\
LaCl$_3$ cb  & 4.95 & 5.26 & 5.82 \\
LaBr$_3 f$   & 3.20 & 3.28 & 3.24 \\
LaBr$_3$ cb  & 3.98 & 4.24 & 4.60 \\
\hline
\end{tabular}
\\
$^*f$-resonance overlaps bottom of conduction band.
\label{gaps}
\end{table}

\begin{figure}
\includegraphics[height=0.9\columnwidth,angle=270]{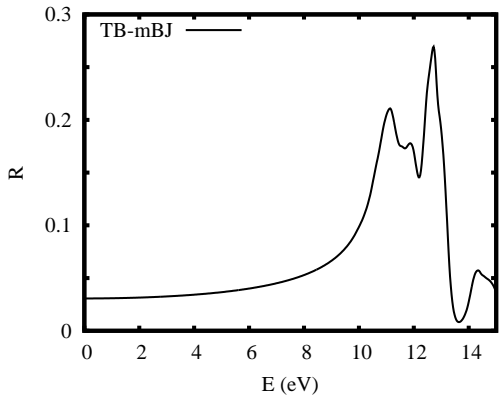}
\caption{Calculated TB-mBJ reflectivity spectrum of BaF$_2$.}
\label{BaF2-refl}
\end{figure}

The experimental band gap of BaF$_2$ is usually
quoted as 11 eV based on the measured
ultraviolet reflectance spectrum. \cite{rubloff}
This spectrum shows several features near the gap: a prominent feature
at 10.00 eV, assigned as an exciton, a weaker feature at 10.57, also
assigned as an exciton, and a direct band edge at 11.0 eV.
Subsequent two photon absorption experiments assigned the gap as
10.6 eV. \cite{tsujibayashi}
The
TB-mBJ direct gap at $X$ of 9.90 eV is $\sim$ 1 eV lower.
Importantly, the experimental spectrum has
three large peaks in the valence band region below
16 eV.
The calculated TB-mBJ reflectance spectrum is shown in Fig. \ref{BaF2-refl}.
There are two prominent peaks that can be matched with the upper two peaks
in the experimental spectrum (Ref. \onlinecite{rubloff})
using a band shift of $\sim$ 1 eV.
This is consistent with the assignment of the lower experimental peak
as excitonic and the conclusion that the TB-mBJ band gap is underestimated
by $\sim$ 1 eV.

\begin{figure}
\includegraphics[height=0.9\columnwidth,angle=270]{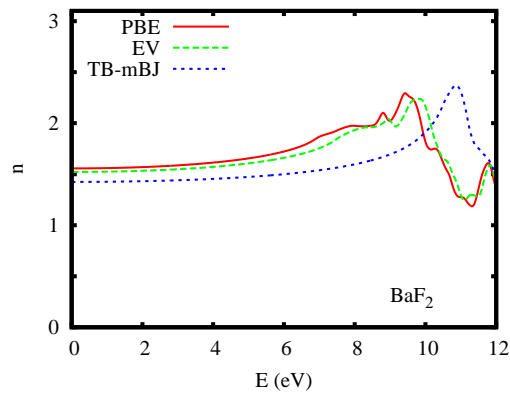}
\caption{(color online)
Calculated refractive index of BaF$_2$ with the three functionals.}
\label{BaF2-refract}
\end{figure}

\begin{figure}
\includegraphics[height=0.9\columnwidth,angle=270]{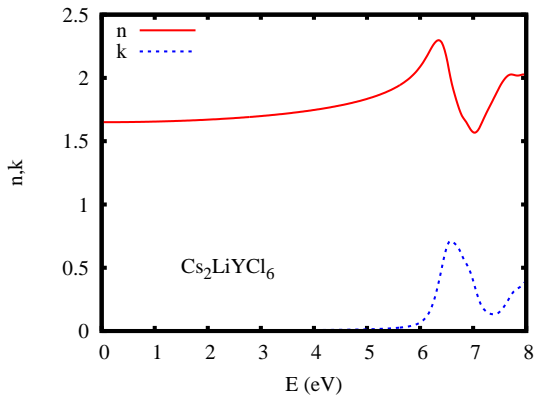}
\caption{(color online) Calculated refractive index of Cs$_2$LiYCl$_6$.}
\label{Cs2LiYCl6-refract}
\end{figure}

The calculated refractive index is shown as a function of energy for the
three functionals in Fig. \ref{BaF2-refract}.
The refractive indices follow the trend in the band gaps, with the result
of the TB-mBJ functional lowest.
Experimentally, the refractive index of BaF$_2$ is practically constant,
rising from $n$=1.465 at low energy to $n_D$=1.474 at 589 nm (2.107 eV).
\cite{shannon}
This is in close agreement with the results using the TB-mBJ functional.
We obtain a low frequency value of $n$=1.42 with a weak dispersion as 
in experiment.
The low frequency PBE value is higher at $n$=1.56 and shows a stronger
energy dependence consistent with the smaller gap, while the
EV value of $n$=1.52 is intermediate.

This result
shows that the TB-mBJ functional not only improves the band gap, but also
improves the optical response of this halide relative to the PBE
functional, which is regarded as state of the art for calculations of
total energies. This is important
because the response depends not only on band energies but also
on the wavefunctions, specifically through the dipole matrix elements.
The charge density and total energy are fundamental quantities within
density functional theory while the single particle energies are not.
It is strongly thought that the Kohn-Sham eigenvalues of the exact
density functional will not produce correct 
band gaps in insulators, even though its charge density and static
response will be exact. \cite{jones}
Therefore it is of interest to note that the improvement in the band
gap of semi-local functionals, particularly going to the EV and then
the TB-mBJ functional improve the static response at the same time as
they improve the band gap.
In the following we focus on results obtained
with the TB-mBJ functional.

The calculated TB-mBJ band gap of 5.97 eV
for the high light output elpasolite scintillator, Cs$_2$LiYCl$_6$
agrees well with the experimental onset of optical absorption at 5.9 eV. 
The refractive index is shown in Fig. \ref{Cs2LiYCl6-refract}.
\cite{loef}
The low frequency limit is $n$=1.65. There is more dispersion than
in BaF$_2$, but the spectrum is otherwise featureless almost up to the
band edge.
We are not aware of experimental measurements of the optical properties
of this material.

The isostructural tri-iodides, YI$_3$ and BiI$_3$ offer an interesting
contrast.
As mentioned, when activated with Ce$^{3+}$, YI$_3$ is a very high light
output scintillator, while BiI$_3$ is not.
The emission of YI$_3$:Ce$^{3+}$ starts at $\sim$ 2.9 eV and
peaks at 2.26 eV. \cite{glodo-yi3}
The high energy
onset is below our calculated TB-mBJ gap of 3.31 eV, but is higher
than the gap obtained with the standard PBE GGA and
prior calculations done with the local density approximation (LDA).
\cite{boutchko}
The difference between YI$_3$ and BiI$_3$ can be understood in terms
of the electronic structure, specifically that the band
gap of BiI$_3$ is too small and that this is due to low lying
Bi $6p$ states.

\begin{figure}
\includegraphics[height=0.9\columnwidth,angle=270]{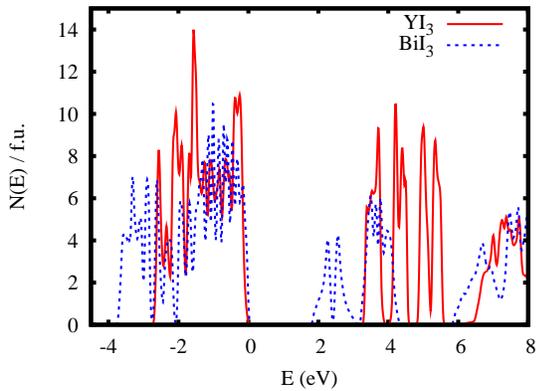}
\caption{(color online)
Comparison of the electronic DOS of YI$_3$ and Bi$_3$ as
obtained with the TB-mBJ functional.}
\label{YI3-Bi3-dos}
\end{figure}

\begin{figure}
\includegraphics[height=0.9\columnwidth,angle=270]{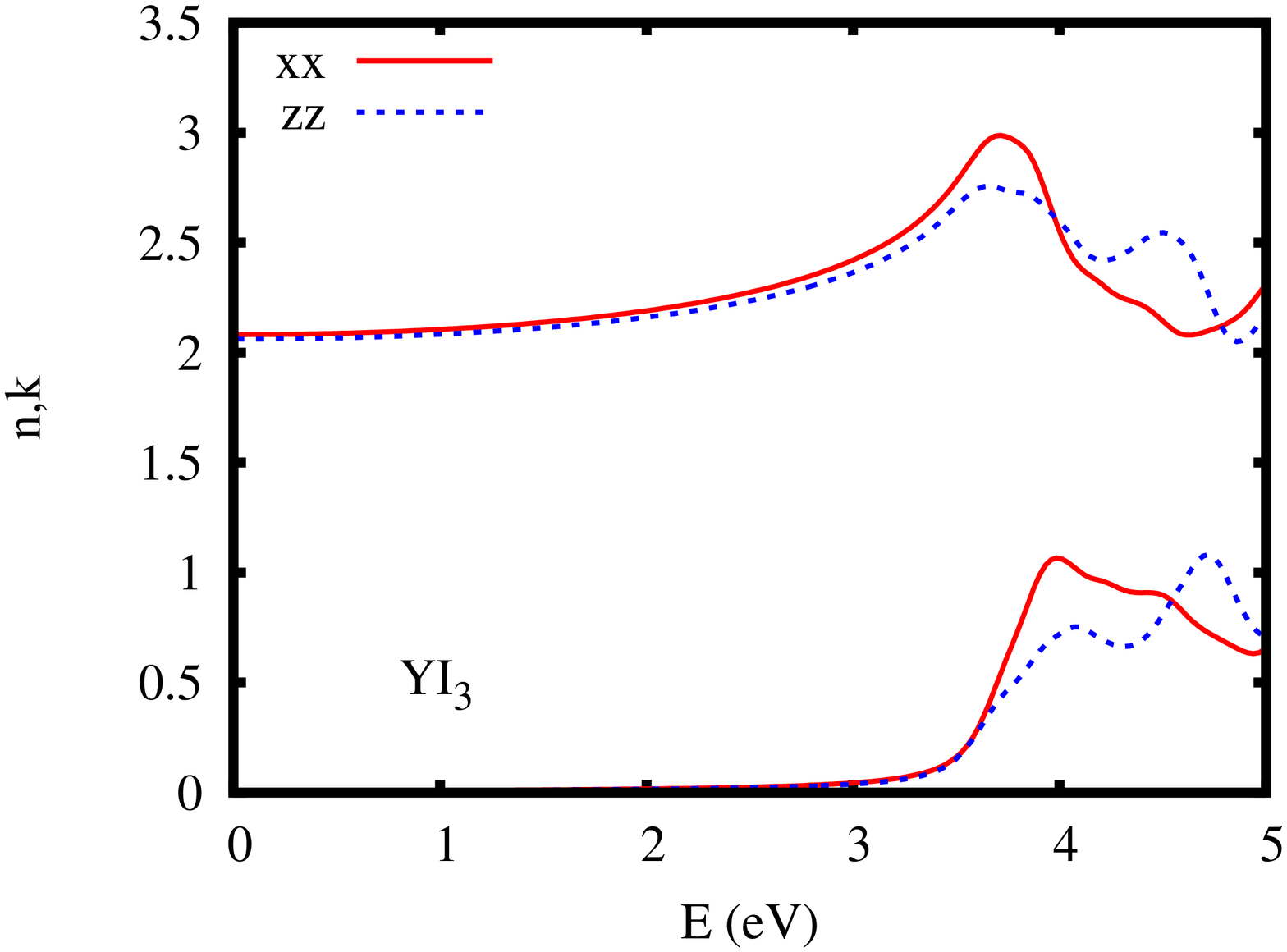}
\includegraphics[height=0.9\columnwidth,angle=270]{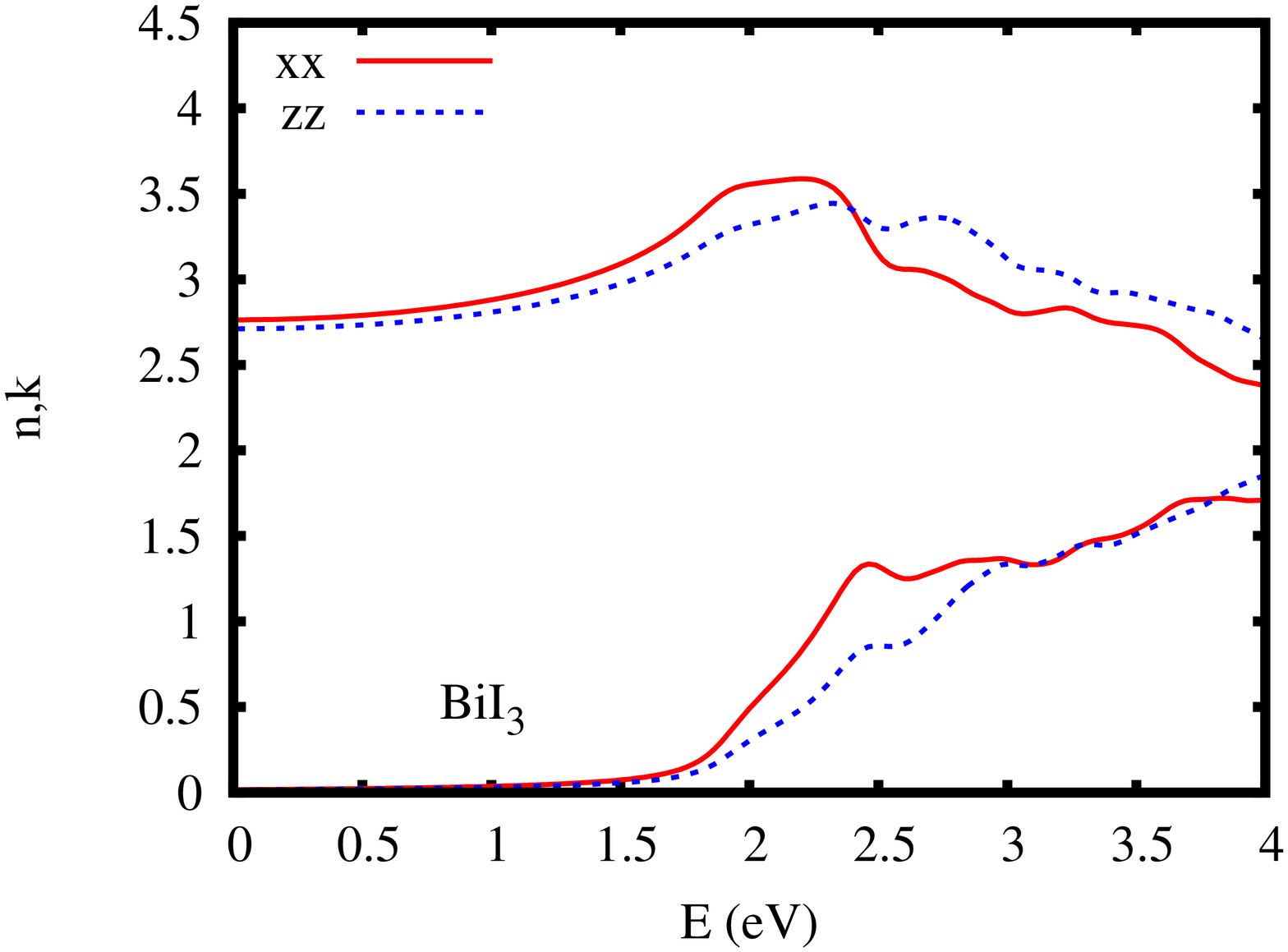}
\caption{(color online)
Calculated refraction of YI$_3$ (top)
and BiI$_3$ (bottom) as a function of energy
based on the TB-mBJ functional.}
\label{YI3-refr}
\end{figure}

\begin{figure}
\includegraphics[width=0.65\columnwidth,angle=0]{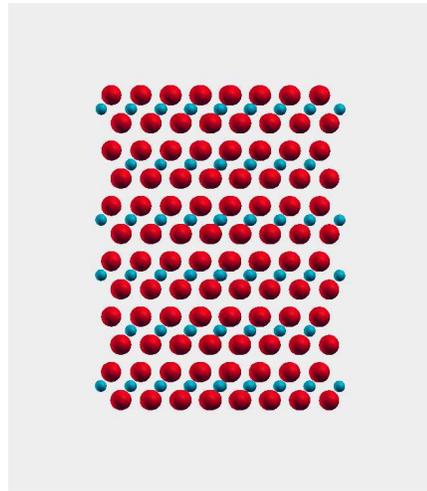}
\caption{(color online)
Crystal structure of YI$_3$ with the relaxed atomic positions.
The large red balls denote I, while the smaller gray balls are Y. Note the
layered structure.}
\label{YI3-struct}
\end{figure}

In order for scintillation to take place an electron hole pair must
recombine at the Ce$^{3+}$ site. For this to be efficient both
the upper and lower states for the activator ion should be in the band
gap. This is evidently not possible in BiI$_3$. As mentioned, the
Ce$^{3+}$ emission
in YI$_3$ has a high energy onset of 2.9 eV, which is similar to that
in other iodides.
This is a lower bound on the energy difference between the upper
and lower states of the activator ion.
A comparison of the calculated TB-mBJ electronic DOS of the two materials
is given in Fig. \ref{YI3-Bi3-dos}.
The lowest conduction bands in YI$_3$ are formed from Y $4d$ states.
In contrast, there is a manifold of Bi $6p$ states making up the conduction
bands in BiI$_3$, and these occur at lower energy. This leads to a 1.49 eV
lower band gap in the Bi compound.
Our TB-mBJ gap for BiI$_3$ is substantially higher than that obtained in
prior calculations using other functionals, \cite{schluter,yorikawa}
but it is still only 1.82 eV. The experimental optical
gap is 2.0 eV. \cite{jellison}
This does not leave a
sufficient energy range for both the upper and lower states to be in the gap.
This is similar to the problem in activating Pb phosphate glasses
with rare earths. \cite{singh-phosphate}
In fact, considering the fact that, all things being
equal, lower gaps favor higher light
output, but that too low a gap prevents activation, the gap of YI$_3$ is
probably close to optimum for a Ce$^{3+}$ activated material.

\begin{figure}
\includegraphics[height=0.9\columnwidth,angle=270]{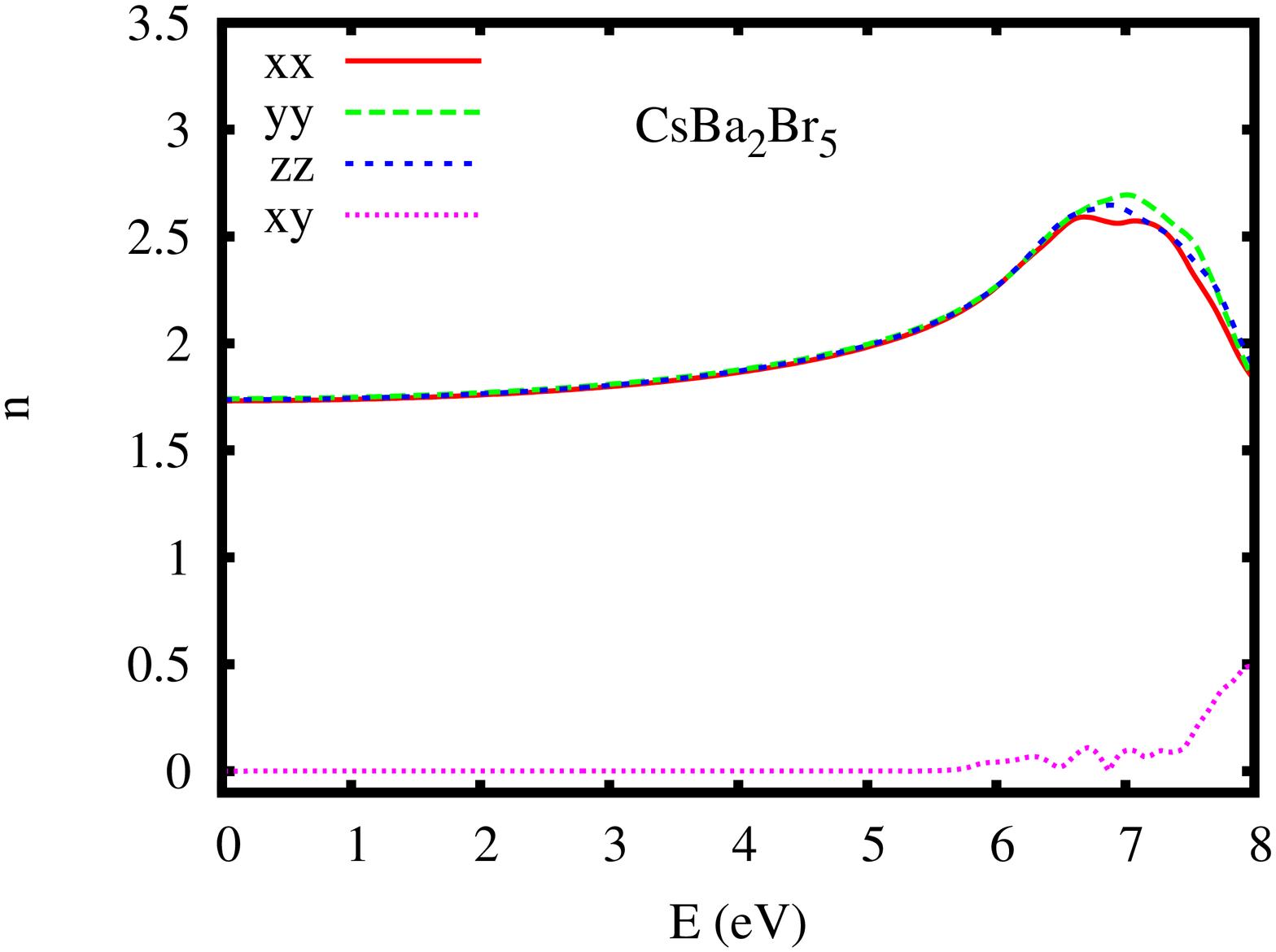}
\includegraphics[height=0.9\columnwidth,angle=270]{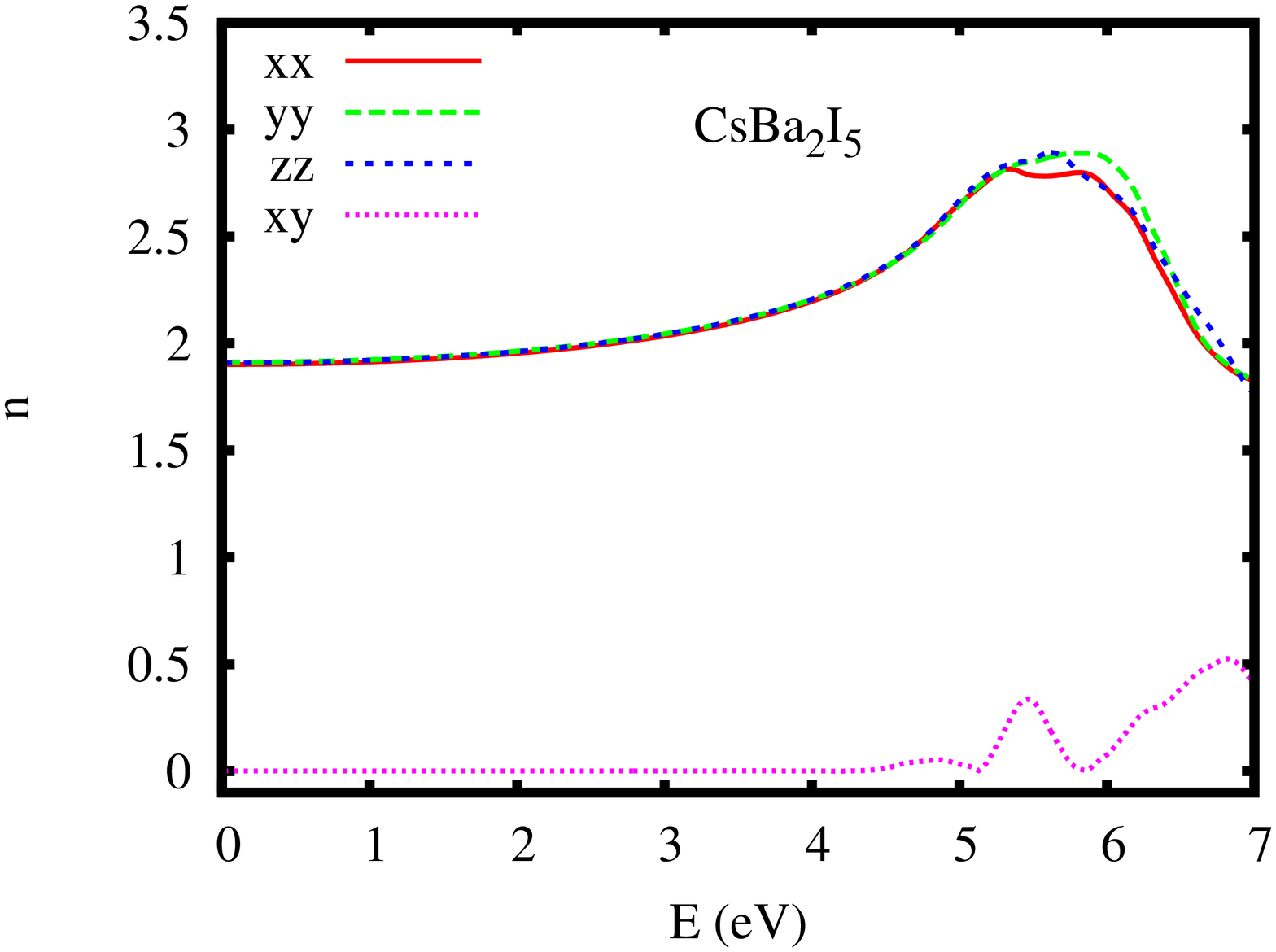}
\caption{(color online)
Calculated refraction for CsBa$_2$Br$_5$ (top)
and CsBa$_2$I$_5$ (bottom) based on the TB-mBJ functional.}
\label{CsBa2Br5-refr}
\end{figure}

\begin{figure}
\includegraphics[height=0.9\columnwidth,angle=270]{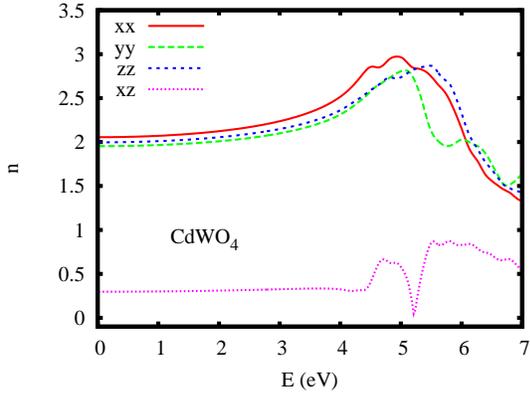}
\caption{(color online)
Calculated refraction for CdWO$_4$ with the TB-mBJ functional.}
\label{CdWO4-refr}
\end{figure}

The refractive indices of YI$_3$ and BiI$_3$ are
presented in Fig. \ref{YI3-refr}.
Surprisingly, the refractive index of YI$_3$ is almost isotropic, even though
the material itself is structurally very anisotropic as shown in Fig.
\ref{YI3-struct}.
The low energy limits of the refractive indices of YI$_3$ are
$n_{xx}$=2.08 and $n_{zz}$=2.06, which is an anisotropy of 1\%.
BiI$_3$ is slightly less isotropic, but is still remarkably so for such
an anisotropic layered crystal structure.
The low energy refractive indices are
$n_{xx}$=2.76 and $n_{zz}$=2.71 for BiI$_3$.
Jellison and co-workers \cite{jellison} reported a value of 3.1 at 1.6 eV.
For comparison, our calculated values at 1.6 eV are 3.16 in-plane ($xx$)
and 3.02
out-of-plane ($zz$), in excellent agreement with the experiment.

We find a similar remarkable near
isotropy of the optical properties in several other halides as well.
This is the case even for monoclinic CsBa$_2$Br$_5$ and CsBa$_2$I$_5$
as shown in Fig. \ref{CsBa2Br5-refr}.
The three diagonal components of the refractive index are practically
the same almost up to the band edge, while the off-diagonal component
associated with the monoclinic symmetry is practically zero over this
energy range.

\begin{figure}
\includegraphics[height=0.9\columnwidth,angle=270]{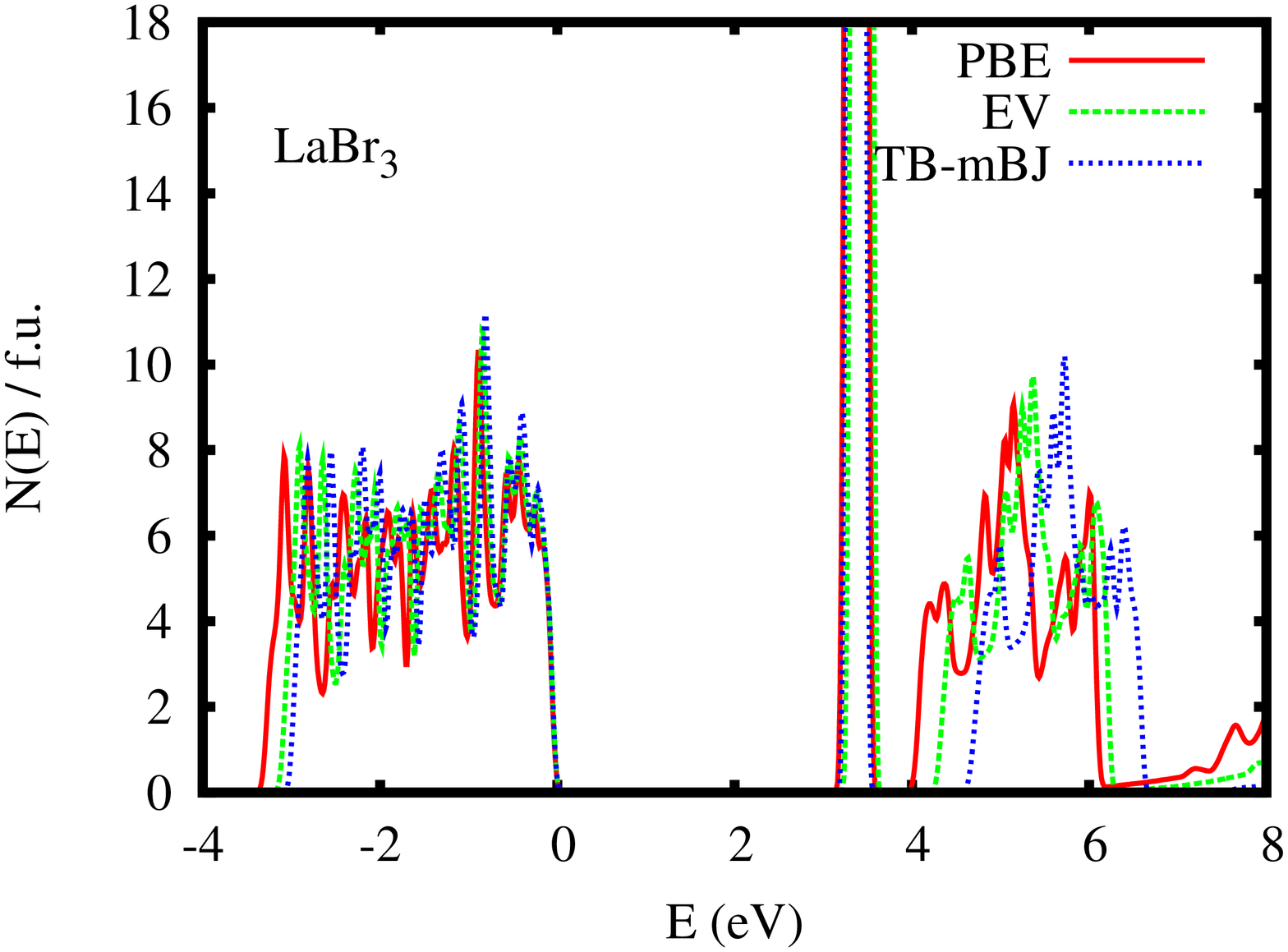}
\includegraphics[height=0.9\columnwidth,angle=270]{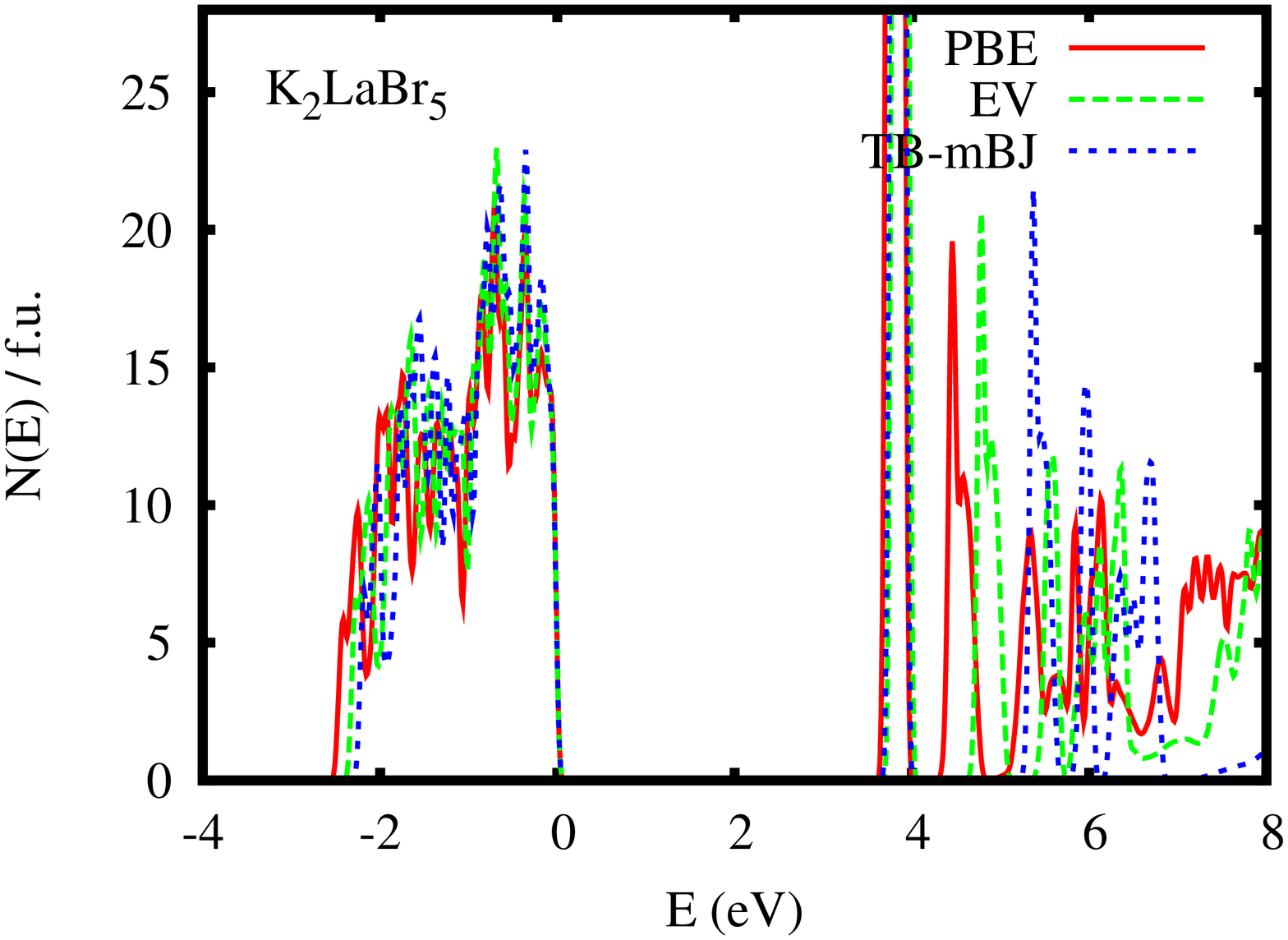}
\caption{(color online)
Comparison of the DOS with the three functionals
for LaBr$_3$ (top) and K$_2$LaBr$_5$ (bottom). Note that the position of
the $f$-resonance that sets the fundamental gap is relatively unaffected
by the changes in functional.}
\label{La-dos}
\end{figure}

For comparison, we also performed calculations for monoclinic CdWO$_4$.
We followed the same procedure relaxing the internal coordinates
using the PBE functional and then performing optical calculations
using the TB-mBJ functional. The calculated band gaps are
2.99 eV with the PBE functional and 4.16 eV with the TB-mBJ form.
The PBE value is similar to that obtained previously by Abraham
and co-workers. \cite{abraham}
The TB-mBJ value is close to the experimental value of 3.8 -- 4.1 eV (see
Ref. \onlinecite{abraham}).
The refractive index is shown in Fig. \ref{CdWO4-refr}.
As may be seen,
in contrast to the halides,
it is quite anisotropic with a substantial off-diagonal
component.

\begin{figure}
\includegraphics[height=0.9\columnwidth,angle=270]{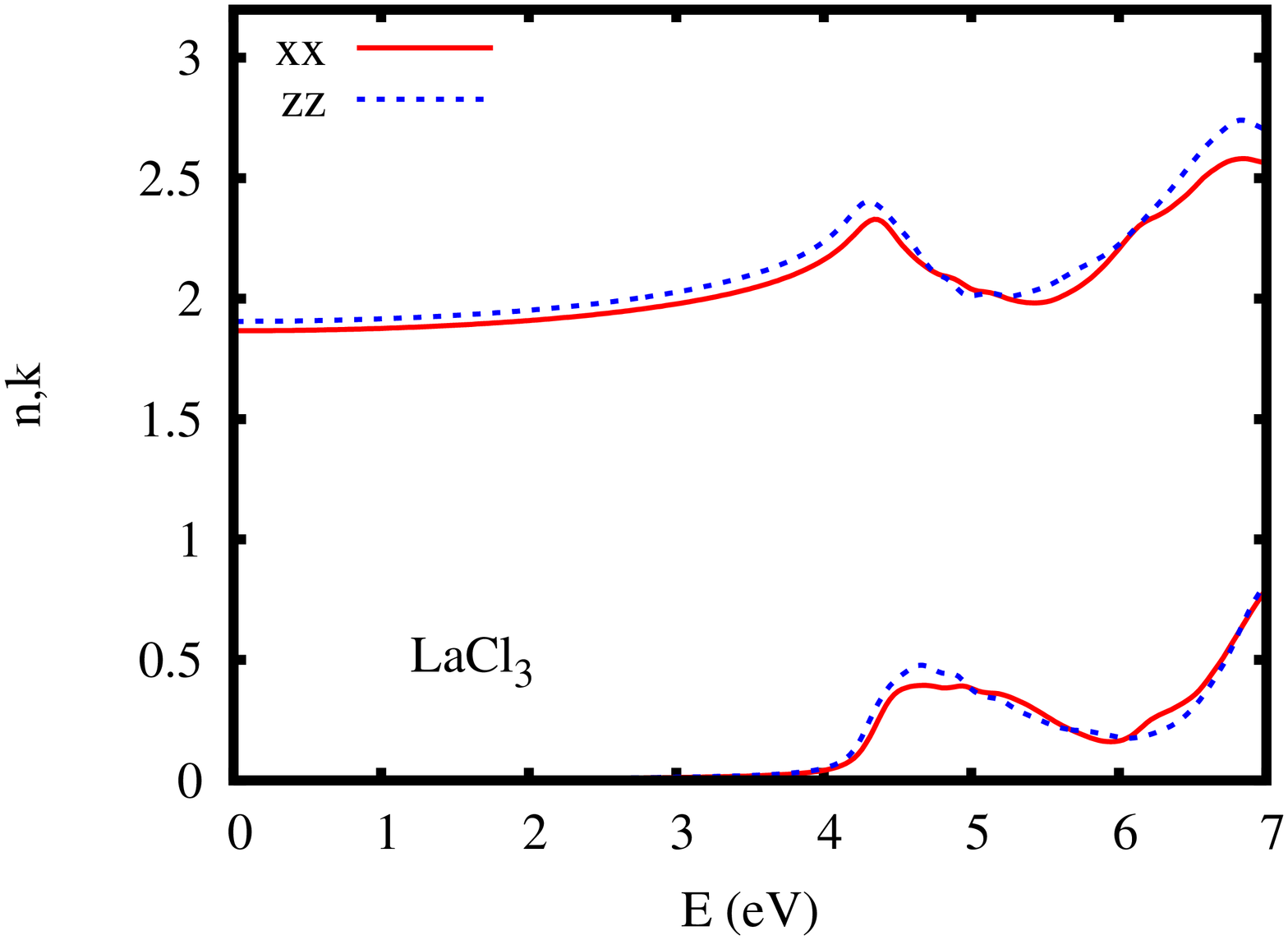}
\includegraphics[height=0.9\columnwidth,angle=270]{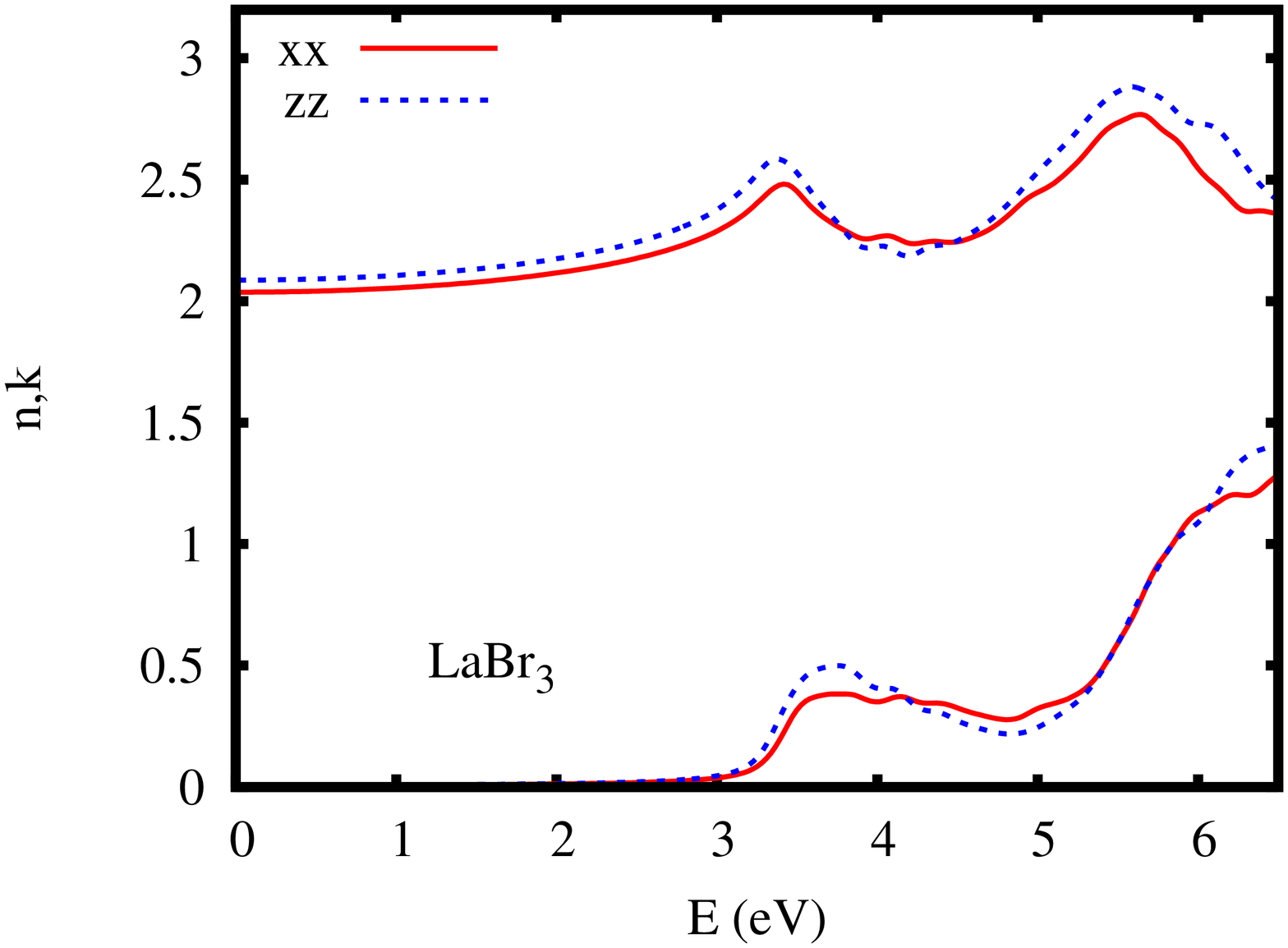}
\caption{(color online)
Calculated refraction for LaCl$_3$(top)
and LaBr$_3$ (bottom) based on the TB-mBJ functional.}
\label{LaCl3-refr}
\end{figure}

\begin{figure}
\includegraphics[height=0.9\columnwidth,angle=270]{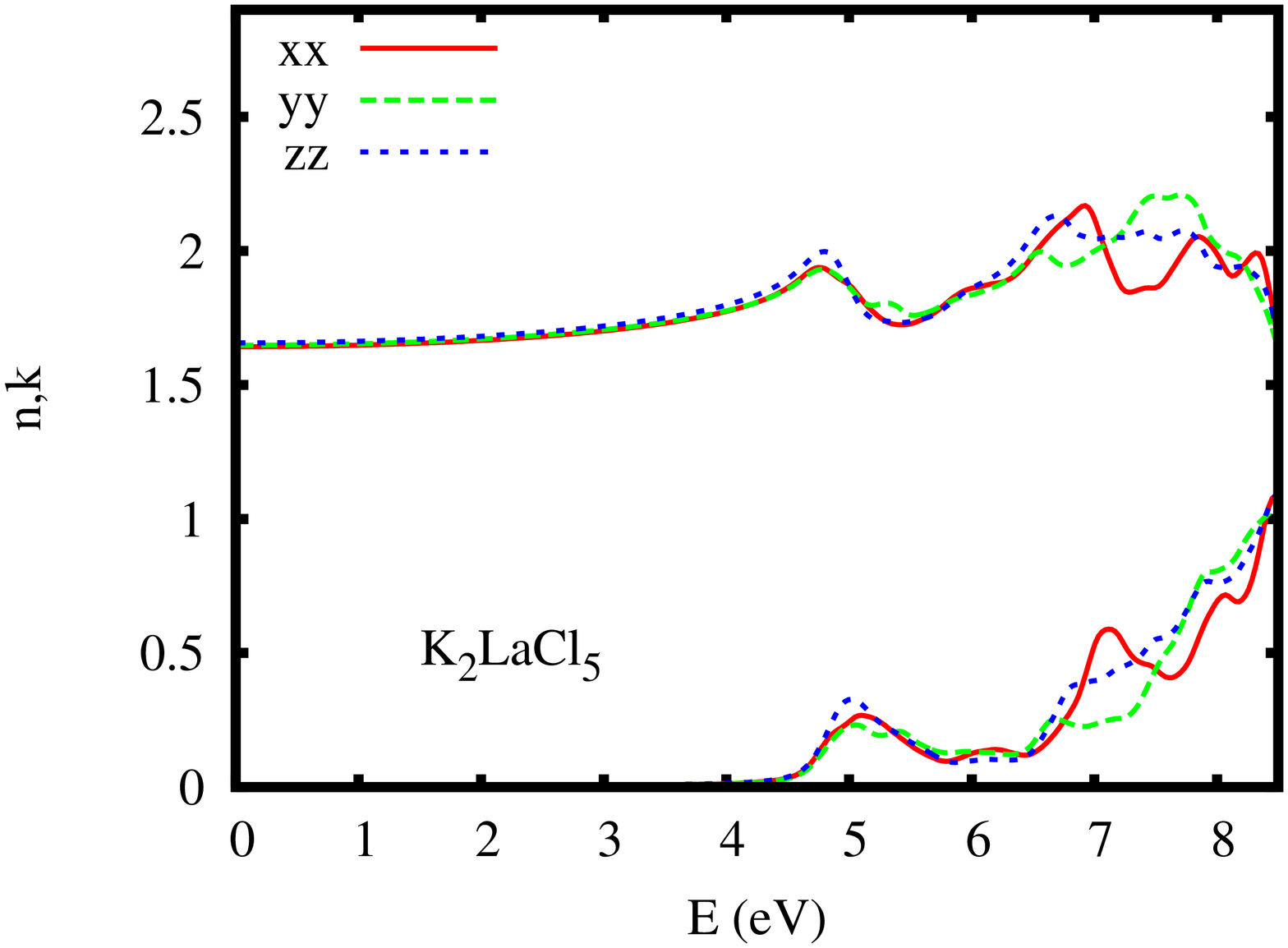}
\includegraphics[height=0.9\columnwidth,angle=270]{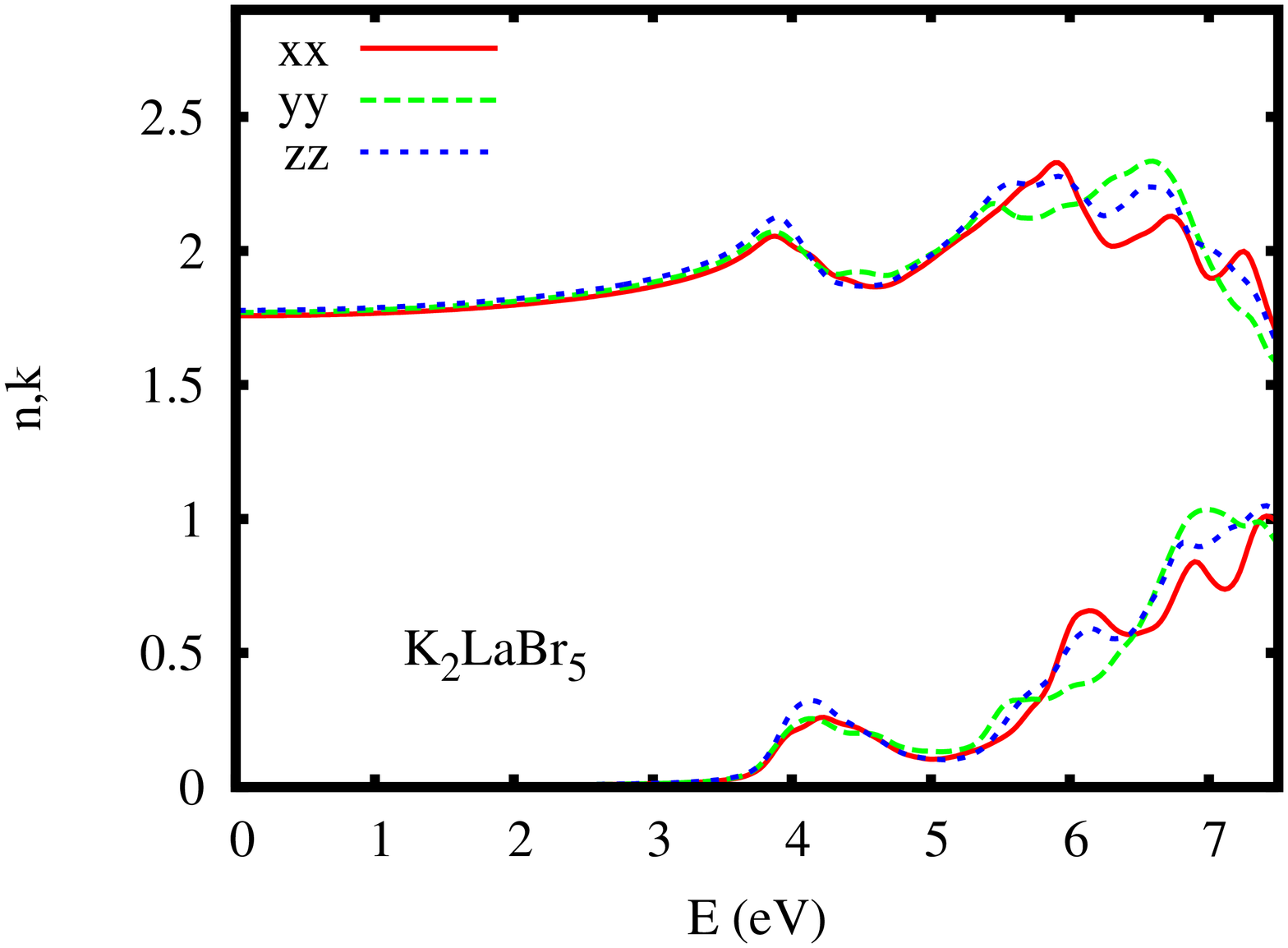}
\includegraphics[height=0.9\columnwidth,angle=270]{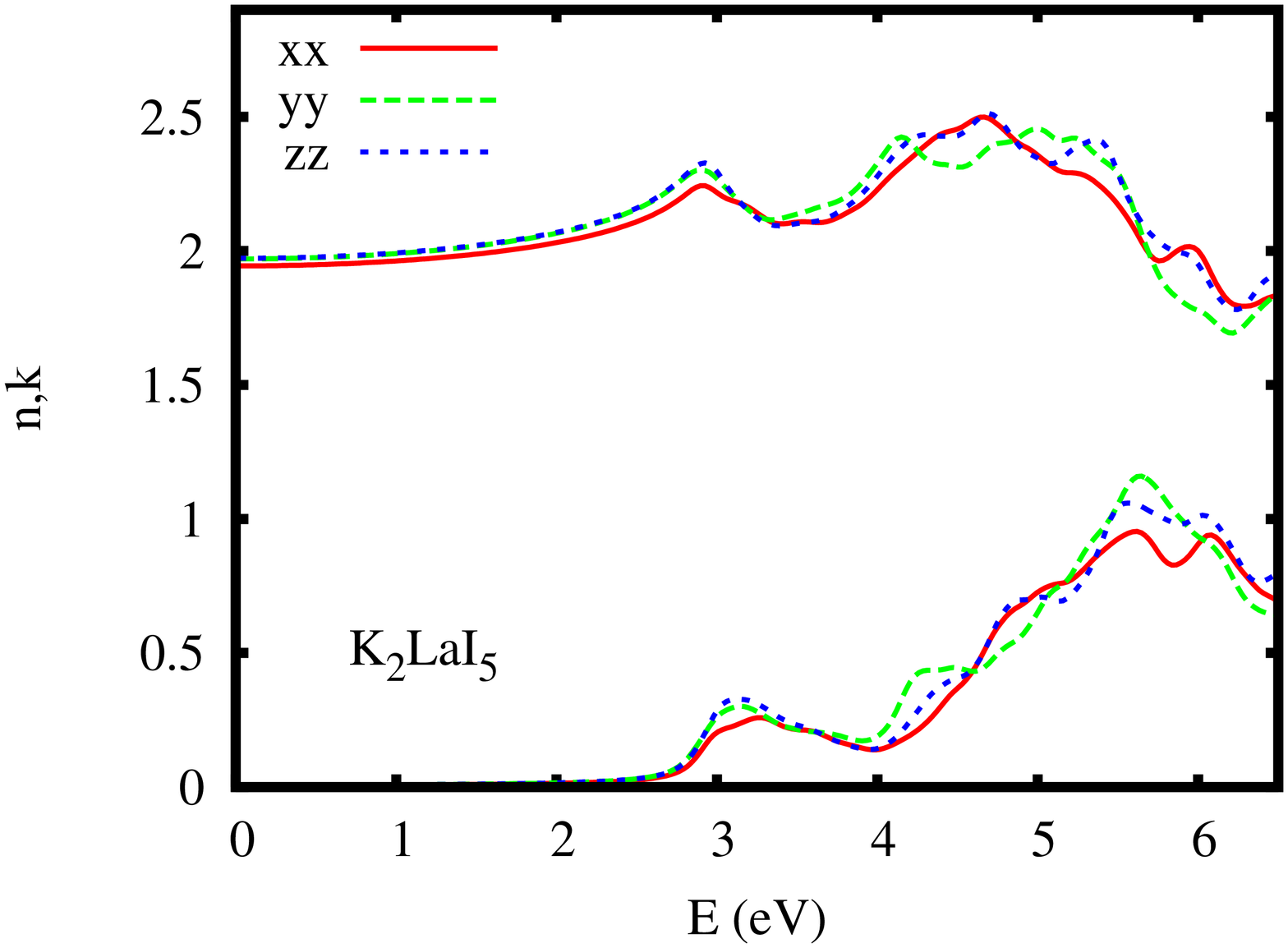}
\caption{(color online)
Calculated refraction for K$_2$LaCl$_5$(top), K$_2$LaBr$_5$ (middle)
and K$_2$LaI$_5$ (bottom) based on the TB-mBJ functional.}
\label{K2LaCl5-refr}
\end{figure}

\begin{figure}
\includegraphics[height=0.9\columnwidth,angle=270]{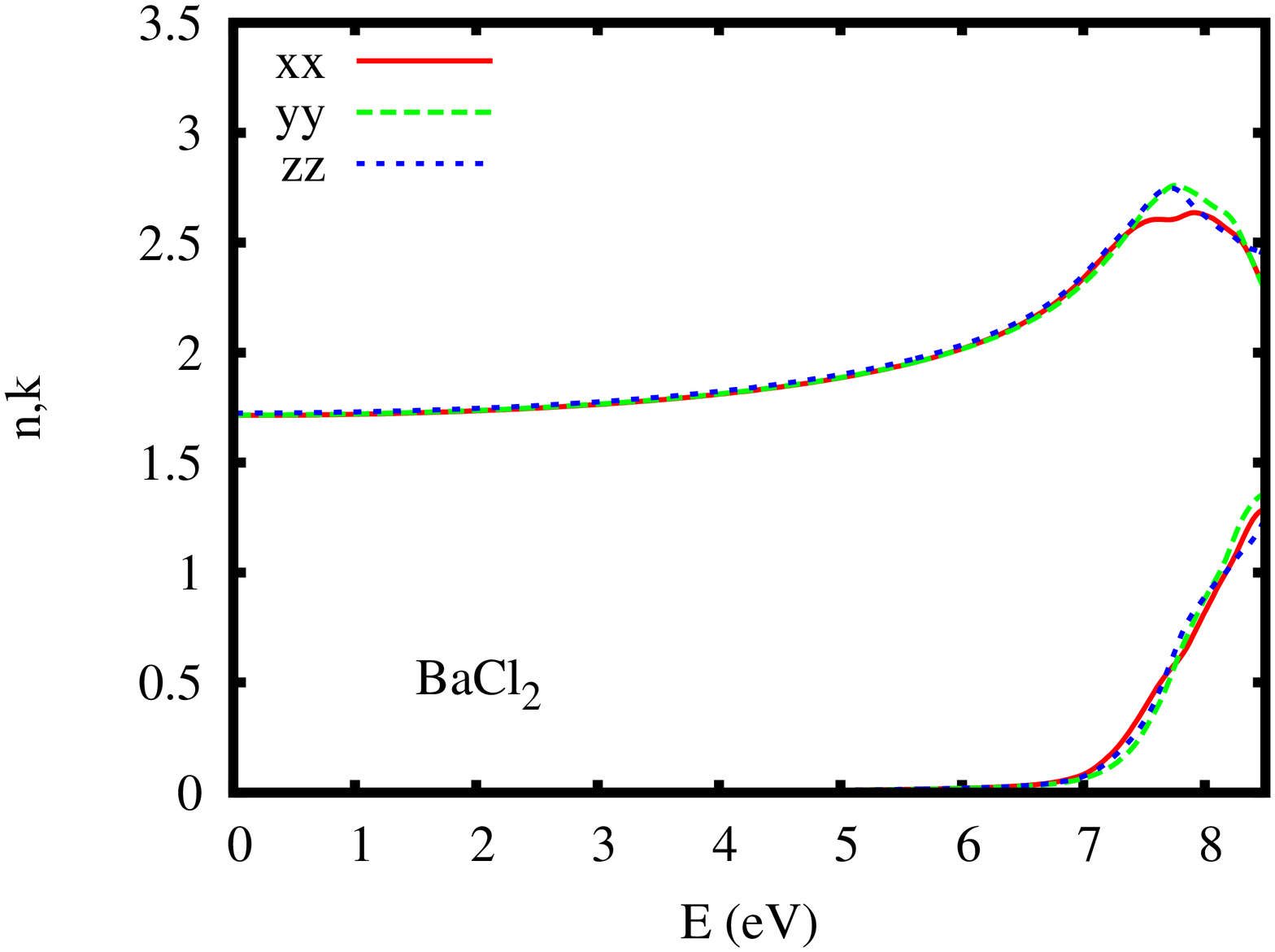}
\includegraphics[height=0.9\columnwidth,angle=270]{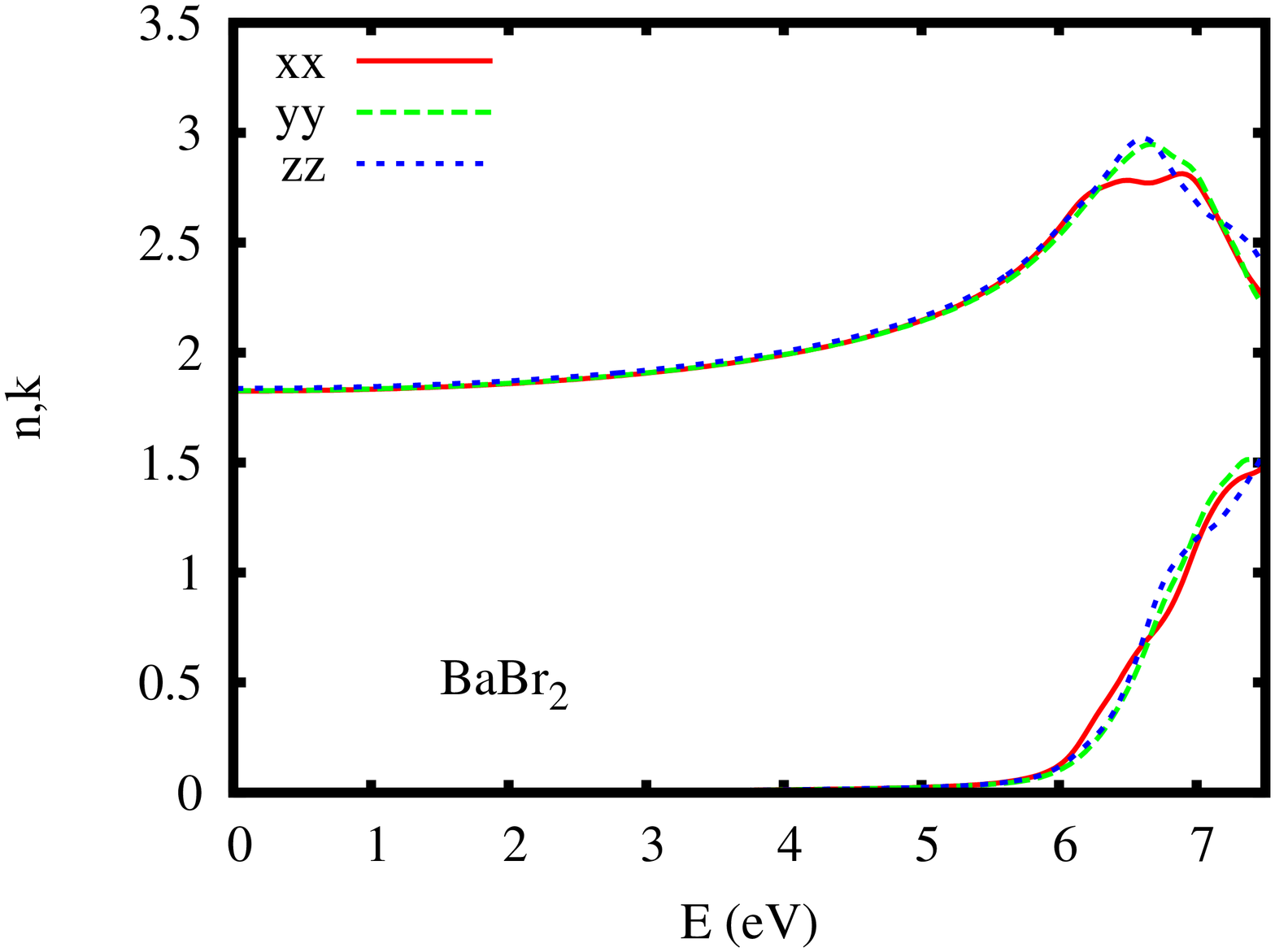}
\includegraphics[height=0.9\columnwidth,angle=270]{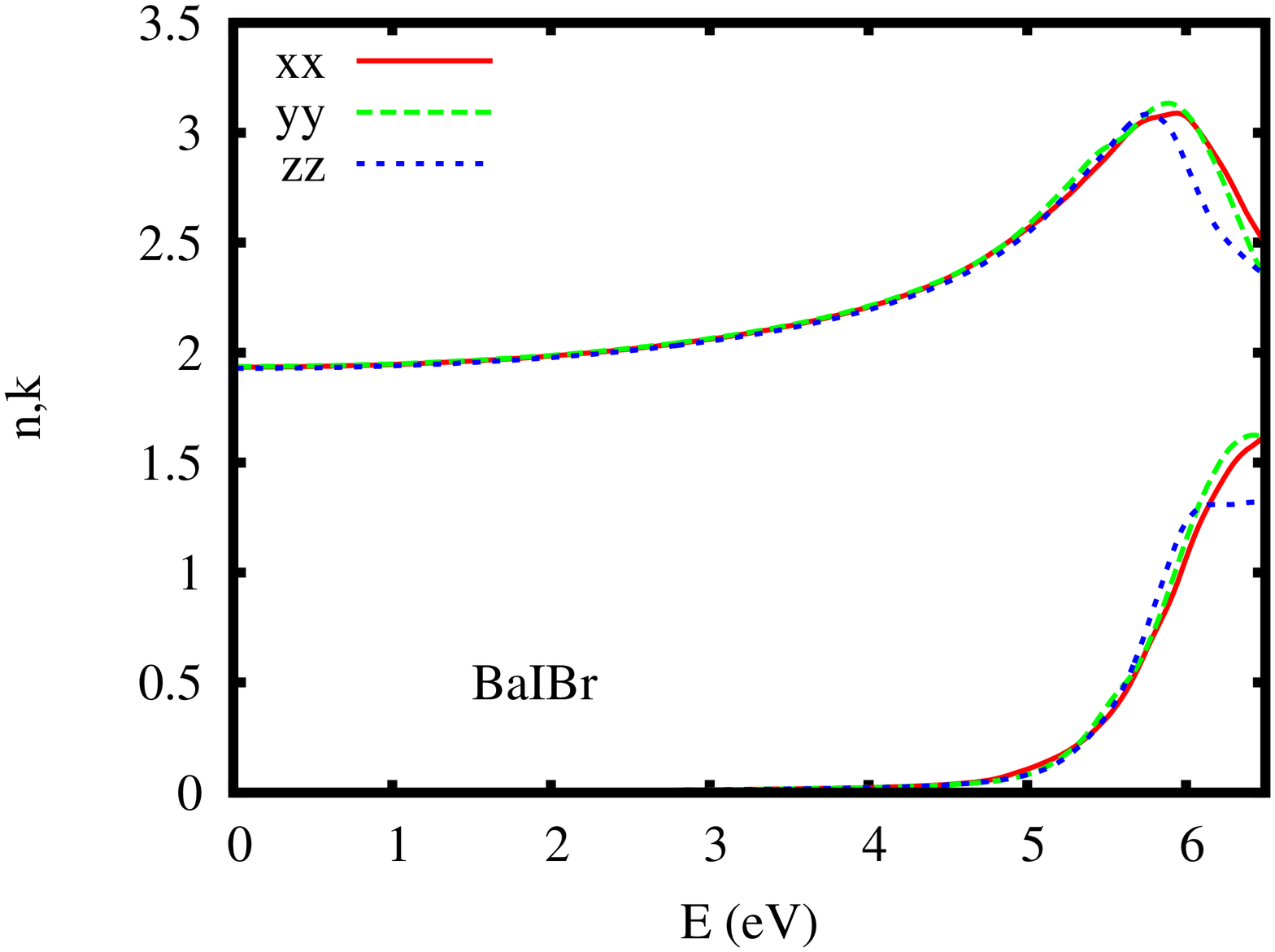}
\includegraphics[height=0.9\columnwidth,angle=270]{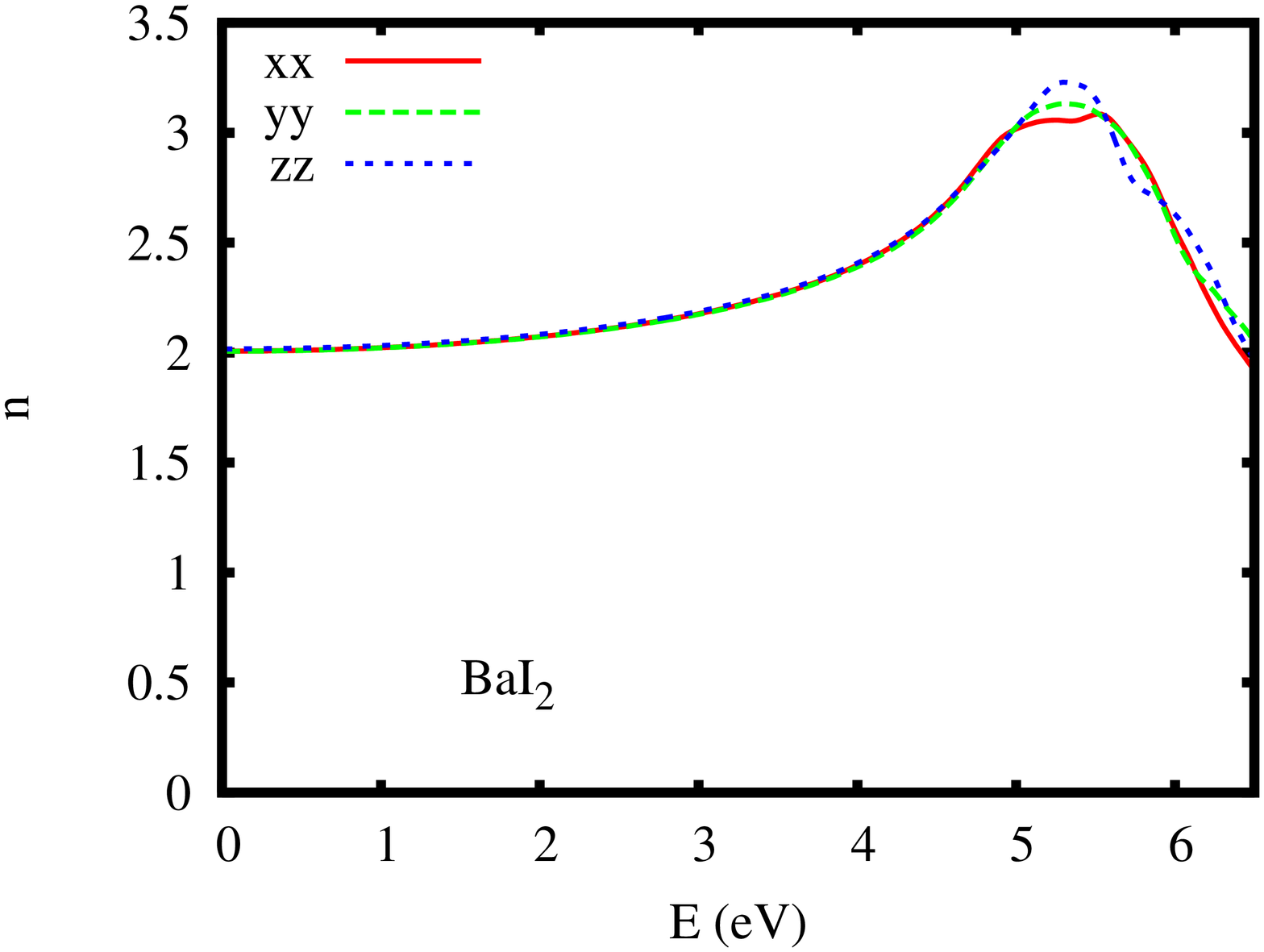}
\caption{(color online)
Calculated refraction for BaCl$_2$, BaBr$_2$, BaIBr and BaI$_2$,
from top to bottom using the TB-mBJ functional.}
\label{BaCl2-refr}
\end{figure}

Returning to halides, we next discuss La containing materials.
The band structure of these materials may be described as
that of an equivalent material based on a simple trivalent ion, e.g. Y,
plus additional unoccupied $f$ derived bands that occur in the band gap.
The $f$ bands comprise the so-called $f$-resonance. 
In Table \ref{gaps} we give both the value of the fundamental gap, which
is from the top of the
halogen $p$ derived valence bands to the bottom of the
$f$-resonance as well as a larger gap, denoted ``cb", which is from the
top of the valence bands to the bottom of the conduction bands, excluding
the $f$-resonance.

\begin{figure}
\includegraphics[height=0.9\columnwidth,angle=270]{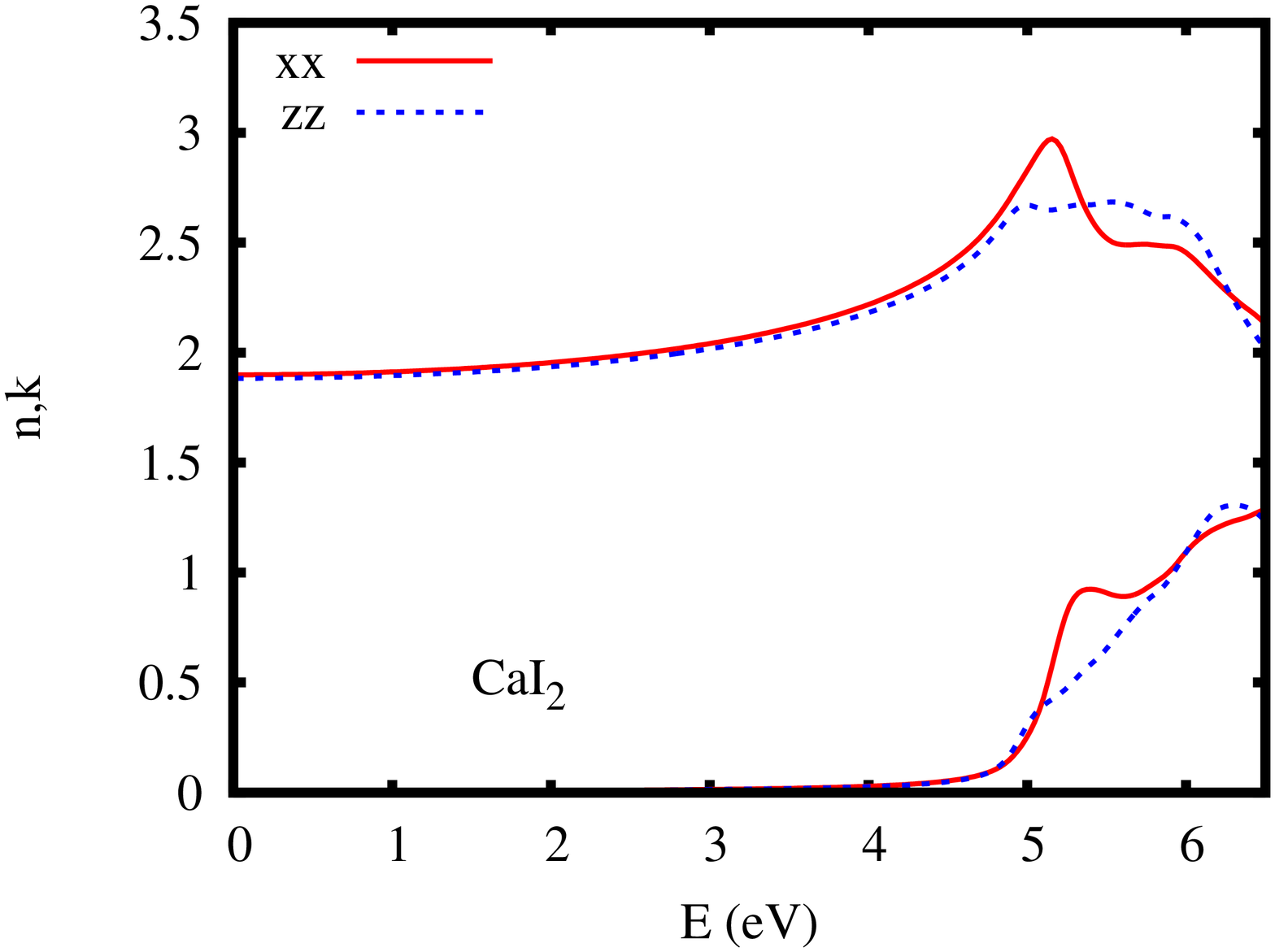}
\caption{(color online)
Calculated refraction for CaI$_2$
based on the TB-mBJ functional.}
\label{CaI2-refr}
\end{figure}

\begin{figure}
\includegraphics[height=0.9\columnwidth,angle=270]{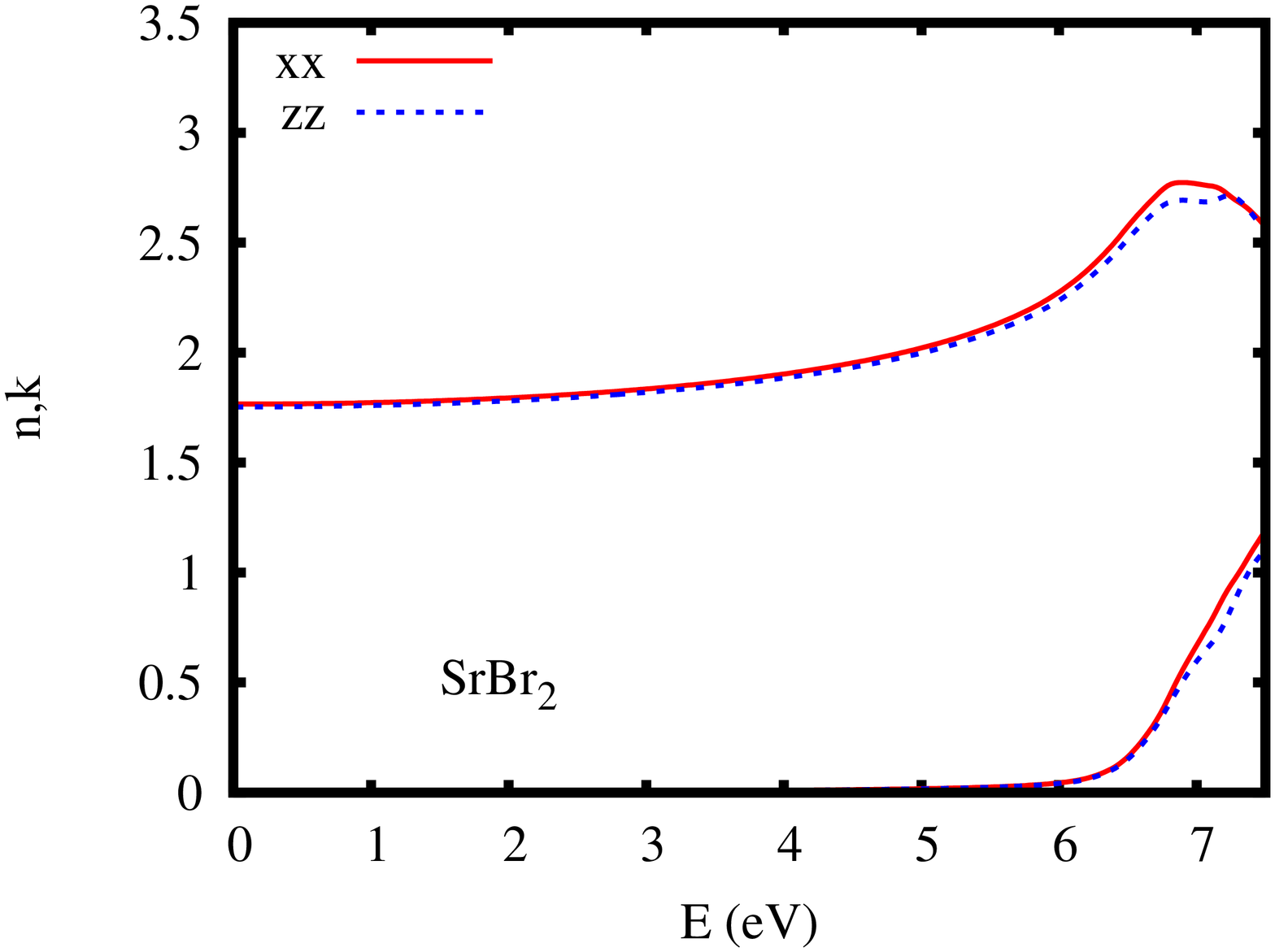}
\caption{(color online)
Calculated refraction for SrBr$_2$
based on the TB-mBJ functional.}
\label{SrBr2-refr}
\end{figure}

We find that in contrast to the band gaps of the other materials studied,
the fundamental gap of these materials is practically unchanged upon
going from the PBE to the TB-mBJ functional. That is the position of the
$f$ bands relative to the valence band maximum is almost the same with these
three functionals.
This is illustrated for LaBr$_3$ and K$_2$LaBr$_5$ in Fig. \ref{La-dos}.
The position of the higher lying non-$f$ conduction bands
is, however, increased with the TB-mBJ functional. This does lower the
optical refractive index, even though the fundamental band gap is unchanged.

The calculated refractive indices of LaCl$_3$ and LaBr$_3$ are shown in
Fig. \ref{LaCl3-refr}.
The low energy values are
$n_{xx}$=1.87 and $n_{zz}$=1.91 for LaCl$_3$ and
$n_{xx}$=2.04 and $n_{zz}$=2.09 for LaBr$_3$.
This follows the expected trend where bromides have higher refractive
index than chlorides.
Again, we find low anisotropy in materials that are structurally anisotropic.
The same trend with halogen atomic number and also very low optical anisotropy
is found in K$_2$LaCl$_5$, K$_2$LaBr$_5$ and K$_2$LaI$_5$,
as shown in Fig. \ref{K2LaCl5-refr}. The calculated TB-mBJ
band gap to the
non-$f$ upper conduction band edges (``cb") for
K$_2$LaCl$_5$, K$_2$LaBr$_5$ and K$_2$LaI$_5$
are 6.32 eV, 5.26 eV and 3.99 eV, respectively, as compared to the
experimental estimates of 6.6 eV, 5.5 eV and 4.5 eV.
\cite{loef-215,loef-215a}

The refractive indices of BaCl$_2$, BaBr$_2$, BaIBr and BaI$_2$ are
given in Fig. \ref{BaCl2-refr}, while those of CaI$_2$ and SrBr$_2$
are given in Figs. \ref{CaI2-refr} and \ref{SrBr2-refr}, respectively.
The calculated TB-mBJ band gap for BaCl$_2$ of 6.45 eV is in reasonable
accord with the estimate from Cl-$K$ x-ray spectroscopy of 7.0 eV.
\cite{sugiura}
As mentioned, BaCl$_2$, BaBr$_2$, BaIBr and BaI$_2$ occur in a complex
orthorhombic $Pnma$ structure. SrBr$_2$ is tetragonal, while CaI$_2$
is rhombohedral.
None of these materials is cubic.
CaI$_2$ activated with Eu$^{2+}$
has been known to be an extremely high light output material since the
1960's. \cite{hofstadter-cai2}
However, this hexagonal material has not been used in applications because
of crystal growth problems due to its platelet growth habit.
Nonetheless, again we find only very small anisotropy in the optical
properties for all of these materials,
similar to what we found previously for SrI$_2$. \cite{singh-sri2}

\section{discussion}

We have two main conclusions,
besides the numerical data, which we hope will be useful in improving
the design of scintillator systems for better light coupling.
First of all, we find that the newly developed TB-mBJ functional
greatly improves both the band gaps and the optical properties
in a broad class of halide materials,
consistent with results reported for other compounds. \cite{mbj}
Considering the computational efficiency of this method, which is
similar to standard density functional methods, we expect that this
method will enable optical characterization of new complex 
halide scintillators and perhaps,
considering that band gap is a key parameter,
more effective theoretical screens
for new scintillators.

Secondly, and quite unexpectedly, we find that a wide variety of halide
scintillators based on Cl, Br and I are practically isotropic from an
optical point of view, even though many of them are highly anisotropic
from the point of view of structure and other properties.
The broad range of materials in which this occurs implies that it is a
general feature of halide chemistry rather than a special coincidence for
certain compounds.

Qualitatively, it may be understood from the coordination environments
and general band structure features. In particular, these materials
have relatively wide band gaps due to the large electronegativity  
difference between the cations and the halogen atoms.
The halogen $p$ derived valence bands are narrow compared to typical
metal oxides. This narrow band width suggests that one can understand
the properties in real space instead of depending essentially on detailed
band dispersions. Additionally, the valence band formation comes at
least largely from direct hoping between the halogen
$p$ orbitals on adjacent sites these materials.
The structures can be described in anion contact
terms. \cite{pauling}
From a structural point of view, the anion lattices of these compounds
are distortions of high symmetry structures, with most of the
anisotropy coming from the cation placement in the intertices of the
anion lattice.
This places the cations in locally highly symmetric
cages based on high nearest neighbor anion coordination numbers.
Since the anion bands are relatively non-dispersive,
the main crystal structure dependence comes from the conduction bands.
Therefore, the small optical anisotropy of these
materials can be rationalized in terms of the
highly symmetric local environments of the cations as far as nearest neighbor
anion coordination is concerned.
Further work to elucidate this is clearly needed.

In any case, this result has important implications for the development
of halide scintillators. Gamma spectroscopy requires high quality
uniform crystals of sufficient size to effectively stop the Gamma rays
in the scintillator volume. This typically requires cm sized crystals,
and many applications benefit from still larger sizes. As a result, crystal
growth is one of the main challenges in the development of new
halide scintillators. The near optical isotropy of these materials, however,
suggests that ceramic scintillators of sufficient size to be useful
for Gamma spectroscopy can be made.
While this will require solution of a number
of problems, for example, the development of methods to produce dense
sintered bodies without contamination in these often hygroscopic,
air sensitive materials,
it may enable the use of low symmetry difficult to grow halide
materials for gamma spectroscopy and other scintillator applications.
Furthermore, ceramic materials are generally lower cost than single crystals,
especially if large volumes of material are needed.

\acknowledgements

Work at ORNL was supported by the Department of Energy, 
Nonproliferation and Verification Research and Development, NA-22.

\bibliography{scint}

\end{document}